\pdfoutput=1
\documentclass[fleqn,usenatbib,useAMS,twocolumn]{mnras}

\usepackage{graphicx}	
\usepackage{amsmath}	
\usepackage{amssymb}	
\usepackage{multicol}        
\usepackage{bm}		
\usepackage{pdflscape}	
\usepackage{epstopdf}
\usepackage[T1]{fontenc}
\usepackage{ae,aecompl}

\newcommand{\ms}[1]{\textcolor{black}{#1}}

\newcommand{\beq}{\begin{equation}}
\newcommand{\beqa}{\begin{eqnarray}}
\newcommand{\eeq}{\end{equation}}
\newcommand{\eeqa}{\end{eqnarray}}

\newcommand{\simgt}{\lower.5ex\hbox{$\; \buildrel > \over \sim \;$}}
\newcommand{\simlt}{\lower.5ex\hbox{$\; \buildrel < \over \sim \;$}}

\newcommand{\bd}[1]{\mbox{\boldmath $#1$}}

\newcommand{\bmtheta}{{\bm{\theta}}}

\newcommand{\sigmacr}{\Sigma_{\rm cr}}
\newcommand{\sigmacri}{\Sigma_{\rm cr}^{-1}}

\def\avrg#1{\left\langle #1 \right\rangle}

\usepackage[T1]{fontenc}
\usepackage{ae,aecompl}

\usepackage{newtxtext,newtxmath}

\title[Mock shape catalogues in HSC]
{
Mock galaxy shape catalogues in the Subaru Hyper Suprime-Cam Survey
}

\author[M. Shirasaki et al.]
{Masato Shirasaki$^{1}$\thanks{E-mail: masato.shirasaki@nao.ac.jp},
Takashi Hamana$^{1}$,
Masahiro Takada$^{2}$,
Ryuichi Takahashi$^{3}$,
\newauthor
and Hironao Miyatake$^{2, 4, 5}$
\\
$^{1}$National Astronomical Observatory of Japan, 
Mitaka, Tokyo 181-8588, Japan \\
$^{2}$Kavli Institute for the Physics and Mathematics of the Universe
(WPI),
The University of Tokyo Institutes for Advanced\\
 Study (UTIAS), The University of Tokyo, 5-1-5 Kashiwanoha, Kashiwa-shi, Chiba, 277-8583, Japan\\
$^{3}$Faculty of Science and Technology, Hirosaki University, 
3 Bunkyo-cho, Hirosaki, Aomori, 036-8561, Japan\\
$^{4}$Institute  for  Advanced  Research,  Nagoya  University,  Nagoya
464-8601, Japan\\
$^{5}$Division of Physics and Astrophysical Science, Graduate School
of Science, Nagoya University, Nagoya 464-8602, Japan
}
\begin{document}

\date{}

\pagerange{\pageref{firstpage}--\pageref{lastpage}} \pubyear{2019}

\maketitle

\label{firstpage}

\begin{abstract}
We use the full-sky ray-tracing weak lensing simulations to 
generate 2268 mock catalogues for the Subaru Hyper Suprime-Cam (HSC) survey first-year shear catalogue. Our mock catalogues take into account various effects as in the real data: the survey footprints, 
inhomogeneous angular distribution of source galaxies,
statistical uncertainties in photometric redshift (photo-$z$) estimate, variations in the lensing weight, and the statistical noise in galaxy shape measurements including  
both intrinsic shapes and the measurement errors. 
We then utilize our mock catalogues to evaluate statistical uncertainties expected in measurements of cosmic shear two-point correlations $\xi_{\pm}$ with tomographic redshift information for the HSC survey.
We develop a quasi-analytical formula for the Gaussian sample variance properly taking into account the number of source pairs in the survey footprints. The standard Gaussian formula significantly overestimates or underestimates the mock results by 50\% level.
We also show that different photo-$z$ catalogues or the six disconnected fields, rather than a consecutive geometry, cause variations in the covariance by $\sim 5\%$.
The mock catalogues enable us to study the chi-square distribution for $\xi_{\pm}$. We find the wider distribution than that naively expected for the distribution with the degrees-of-freedom of data vector used. Finally, we propose a method to include non-zero multiplicative bias in mock shape catalogue and show the non-zero multiplicative bias can change the effective shape noise in cosmic shear analyses. Our results suggest an importance of estimating an accurate form of the likelihood function (and therefore the covariance)
for robust cosmological parameter inference from the precise measurements.
\end{abstract}


\section{INTRODUCTION}

Weak gravitational lensing is one of the main subjects in modern cosmology 
to improve our understanding of the universe at low redshifts.
Gravitational lensing analysis has a distinct advantage over other probes, enabling to observe total matter density distributions along a line of sight in an unbiased way, regardless of states of matter density. 
Hence, statistical analyses of gravitational lensing data can provide a complete map of large-scale structures 
in the universe and rich cosmological information printed in the lensing data will reveal 
the nature of dark matter \citep[e.g.][]{2009PhRvD..79b3520I, 2011JCAP...01..022M,2014JHEP...06..162K} and the origin of cosmic acceleration \citep[e.g.][]{2013PhR...530...87W,
2017MNRAS.466.2402S}.
An attractive feature of weak lensing analyses leads to keen competition among different groups.
Ongoing scientific programs include 
the Kilo-Degree Survey (KiDS\footnote{\url{http://kids.strw.leidenuniv.nl/index.php}}), 
the Dark Energy Survey (DES\footnote{\url{https://www.darkenergysurvey.org/}}),
and the Subaru Hyper Suprime-Cam Survey (HSC\footnote{\url{http://hsc.mtk.nao.ac.jp/ssp/}}),
while the Wide Field Infrared Survey Telescope (WFIRST\footnote{\url{https://wfirst.gsfc.nasa.gov/}}),
the Large Synoptic Survey Telescope (LSST\footnote{\url{https://www.lsst.org/}}),
and Euclid\footnote{\url{http://sci.esa.int/euclid/}} will aim at 
measuring the gravitational lensing effect on distant galaxies with higher precision than the current surveys
by covering a wider sky coverage and collecting fainter source galaxies.

As rich lensing data set becomes available and analysis method becomes diverse,
robust estimation of statistical uncertainty in the analyses gains importance \citep[e.g.][]{Hartlap2007, 2013MNRAS.432.1928T, 2016MNRAS.456.2662F, 2017MNRAS.464.4658S, 2018MNRAS.473.4150F} and 
a number of systematic effects must be addressed \citep{2013MNRAS.429..661M, 2017arXiv171003235M}.
A possible approach to overcome these challenges is to carry out numerical simulations 
which capture the relevant gravitational physics for lensing analyses
and create a mock catalogue of lensing data by including various observational effects with appropriate recipe.

\citet{2012MNRAS.427..146H} have utilized 185 independent realizations of CLONE catalogue \citep{2012MNRAS.426.1262H}
and produced the mock shape catalogues of the Canada-France-Hawaii Lensing Survey (CFHTLenS).
The CFHTLenS mock catalogues have a high angular resolution of 0.2 arcmin 
but also suffer from small sky coverages.
It is worth noting single CFHTLenS mock can cover 12.84 square degrees and 
the robust estimation of cosmic shear covariance is limited in the angular range of $0.5-40$ arcmin.
\citet{2017MNRAS.465.1454H} have improved the production of mock catalogues 
for first 450 squared degrees of KiDS (KiDS450) data compared to the case in CFHTLenS by increasing 
the simulation volume and the number of independent realizations \citep{2015MNRAS.450.2857H}.
\ms{The KiDS450 mock catalogues have been produced from 1025 independent set of $N$-body simulations with the box size of 
$505\, h^{-1}\,{\rm Mpc}$ and the sky coverage of each mock realization is set to be 100 squared degrees. 
\citep[see][for details]{2018MNRAS.481.1337H}.}
Note that either of CFHTLenS and KiDS450 mock catalogue is not designed to cover the survey footprint in a single realization.

\citet{2016PhRvD..94b2002B} have considered the mock catalogue production by performing ray-tracing simulation
on a curved sky \citep{2013MNRAS.435..115B} and the resulting mock catalogue for DES Science Verification (DES SV) data 
can cover whole survey footprint of DES SV data (139 squared degrees).
Nevertheless, the number of independent mock catalogues is only 126 since the mock catalogues are based on single full-sky lensing simulation. Hence, direct covariance estimation from mock catalogues seems very challenging with DES SV mocks alone.
The DES collaboration recently updated their weak lensing analyses with DES Year 1 (DES Y1) data \citep{2018PhRvD..98d3528T}.
The fiducial pipeline to evaluate the statistical uncertainty of cosmological analyses with DES Y1 data 
relies on the analytic approach developed in \citet{2014MNRAS.440.1379E, 2017MNRAS.470.2100K, 2017arXiv170609359K}. 
They validated their analytic approach by using mock catalogues of DES Y1 data, while
the DES Y1 mock catalogue is based on log-normal simulation which is 
the approximated version of $N$-body simulation. 
It is uncertain if the approach developed in DES Y1 analyses works in other lensing surveys
with a high source number density, such as the Subaru HSC \citep{2018PASJ...70S..25M},
since accurate modeling of non-linear gravitational growth becomes essential as source number density increases.

This paper presents a set of mock catalogues for the first data of Subaru HSC lensing survey, referred to as HSC S16A.
The HSC S16A mock catalogues are based on 108 quasi-independent full-sky simulations \citep{2017ApJ...850...24T}.
Making the best use of full-sky coverage, we extract six separated fields in HSC S16A survey footprint 
from a full-sky simulation multiple times. In the end, we produce 2268 realizations of HSC S16A mock catalogue with the same sky coverage as the actual shape catalogue. We apply the method in \citet{2017MNRAS.470.3476S}
to the HSC S16A data when producing individual mock catalogues. Our pipeline in mock catalogue production
refers both of simulated lensing data and observed galaxy images on object-by-object basis.
Hence, our mock catalogues have the exact same features appeared in the real catalogue,  
such as the angular position, the posterior distribution of photometric redshift, 
and the lensing weight of each source galaxy.
According to the above features, our mock catalogues can be applied to any cosmic shear analyses with the HSC S16A shape catalogue
and will be useful to evaluate the statistical uncertainty and validate the analysis pipeline.
In this paper, we validate our mock catalogues by measuring two-point correlation functions of galaxy shapes $\xi_{\pm}$ 
and comparing with the mock covariance of $\xi_{\pm}$ and a theoretical prediction in detail.
\ms{Note that the mock catalogs in this paper will be used in our forthcoming cosmological studies with HSC S16A, including 
cosmic shear analysis on real space (Hamana et al. in prep) and galaxy-galaxy lensing analysis (Miyatake et al. in prep)}

This paper is organized as follows.
In Section~\ref{sec:WL}, we introduce the basics of gravitational lensing in galaxy imaging surveys
and a theoretical framework to compute the expectation value of $\xi_{\pm}$ and its statistical uncertainty.
We then describe the HSC S16A shape catalogue which is used for cosmological analyses in first HSC data release in
Section~\ref{sec:data}. The production procedure of the HSC S16A mock catalogues is summarized in Section~\ref{sec:sim}.
In Section~\ref{sec:res}, we compare the average $\xi_{\pm}$ and the covariance of $\xi_{\pm}$ at four different redshifts
measured in 2268 mock catalogues with their theoretical predictions. Apart from detailed comparisons with mock results 
and analytic models, we also study the impact of field variance in the HSC S16A footprint
and methodological difference in source redshift estimation on the mock covariance of $\xi_{\pm}$.
In addition, we quantify the likelihood function of observed $\xi_{\pm}$ in the HSC S16A analysis
by using 2268 mock observations.
Furthermore, we introduce a method to include non-zero mulitiplicative bias in our HSC S16A mock catalogues
in an analysis-dependent way. We conclude this paper in Section~\ref{sec:conclusions}.

\section{WEAK GRAVITATIONAL LENSING}\label{sec:WL}
\subsection{Basics}

We first summarize the basics of gravitational lensing 
induced by large-scale structure.
Weak gravitational lensing effect is usually characterized by
the distortion of image of a source object by the 
following 2D matrix:
\beq
A_{ij} = \frac{\partial \beta^{i}}{\partial \theta^{j}}
           \equiv \left(
\begin{array}{cc}
1-\kappa -\gamma_{1} & -\gamma_{2}  \\
-\gamma_{2} & 1-\kappa+\gamma_{1} \\
\end{array}
\right), \label{distortion_tensor}
\eeq
where $\bd{\theta}$ represents the observed position of a source object,
$\bd{\beta}$ is the true position, 
$\kappa$ is the convergence, and $\gamma$ is the shear.
In the weak lensing regime (i.e., $\kappa, \gamma \ll 1$), 
each component of $A_{ij}$ can be related to
the second derivative of the gravitational potential $\Phi$
\citep[]{Bartelmann2001}.
Using the Poisson equation and the Born approximation, 
one can express the weak lensing convergence field as the weighted 
integral of matter overdensity field $\delta_{\rm m}(\bd{x})$:
\beq
\kappa(\bd{\theta})
= \int_{0}^{\chi_{H}} {\rm d}\chi \ q(\chi)\delta_{\rm m}(\chi,r(\chi)\bd{\theta}), \label{eq:kappa_delta}
\eeq
where $\chi$ is the comoving distance, 
$\chi_{H}$ is the comoving distance up to $z\rightarrow \infty$
and $q(\chi)$ is called lensing kernel.
For a given redshift distribution of source galaxies,
the lensing kernel is expressed as
\beq
q(\chi) = \frac{3}{2}\left( \frac{H_{0}}{c}\right)^2 \Omega_{\rm m0} \frac{r(\chi)}{a(\chi)}\, \int_{\chi}^{\chi_{H}} {\rm d}\chi^{\prime} p(\chi^{\prime})\frac{r(\chi^{\prime}-\chi)}{r(\chi^{\prime})}, \label{eq:lens_kernel}
\eeq
where $r(\chi)$ is the angular diameter distance
and $p(\chi)$ represents the redshift distribution of source galaxies
normalized to \ms{$\int_{0}^{\chi_H} {\rm d}\chi \, p(\chi) =1$}.

Power spectrum is a relevant statistical quantity in two-point correlation analysis in weak lensing surveys.
For two source galaxy populations with different $p(\chi)$, it is defined as
\beq
\langle \tilde{\kappa}_{a}(\bd{\ell}_{1}) \tilde{\kappa}_{b}(\bd{\ell}_{2})\rangle
\equiv (2\pi)^2 \, \delta^{(2)}(\bd{\ell}_{1}-\bd{\ell}_{2})\, P_{\kappa, ab}(\ell),
\eeq
where 
$\tilde{\kappa}$ is the Fourier counterpart of convergence,
the indices $a$ and $b$ represent any two source populations,
$\delta^{(n)}(\bd{x})$ is the Dirac delta function in $n$-dimensional space,
and $P_{\kappa,ab}$ is the convergence power spectrum. 
Under the Limber approximation \citep{Limber:1954zz}, we can express the power spectrum as
\beq
P_{\kappa, ab}(\ell) = \int_{0}^{\chi_H}\, \frac{\rm d\chi}{r^2(\chi)}\, q_{a}(\chi)\, q_{b}(\chi) \, P_{\rm m}\left(\frac{\ell}{r(\chi)}, z(\chi)\right),
\label{eq:P_kappa}
\eeq
where $q_{a}$ is the lensing kernel for the source catalogue ``$a$" and it is defined as in Eq.~(\ref{eq:lens_kernel}).
In Eq.~(\ref{eq:P_kappa}), $P_{\rm m}(k,z)$ represents the non-linear matter power spectrum for wavelength of $k$ at redshift $z$.

\subsection{Cosmic shear}
\label{subsec:cosmic_shear}

\ms{The observed galaxy shape is commonly used as an estimator of weak-lensing shear $\gamma$.}
\ms{In this paper, we use distortion of galaxy as an estimator, and denote it as $\epsilon$.}
\ms{The definition of $\epsilon$ is provided in Eq.~(\ref{eq:def_epsilon}).}
In wide-area galaxy imaging surveys, 
a correlation function method is most conventionally
used for cosmological analysis (e.g. \citealp{2017MNRAS.465.1454H}, \citealp{2017MNRAS.471.4412K}, \citealp{2018MNRAS.476.4662V}, \citealp{2018PhRvD..98d3528T}, \citealp{2018arXiv180909148H} and see \citealp{2017MNRAS.471.4412K}, \citealp{2018MNRAS.476.4662V} for some exceptions).
One can estimate the correlation function by cross-correlating lensing shear of two galaxies as a function of separation angles:
\beq
\xi_{\pm}(\theta) = 
\left.\left\langle 
\gamma_{t}(\bd{\theta}_1)\gamma_{t}(\bd{\theta}_2) 
\pm
\gamma_{\times}(\bd{\theta}_1)\gamma_{\times}(\bd{\theta}_2) 
\right\rangle\right|_{|\bd{\theta}_1-\bd{\theta}_2|=\theta}~, \label{eq:xi_pm}
\eeq
where the average is done over all pairs of galaxies separated by a fixed separation angle, $\theta$ in the above case.
In Eq.~(\ref{eq:xi_pm}), we define the tangential and cross components of 
$\gamma=\gamma_{1}+i\gamma_{2}$ as
\beq
\gamma_{t} \equiv -{\it Re}\left[\gamma e^{-2i\phi}\right], 
\, \, \, \, \, \,
\gamma_{\times} \equiv -{\it Im}\left[\gamma e^{-2i\phi}\right],
\eeq
where $\phi$ represents the polar angle of separation vector between two galaxies from the 1st axis, \bd{\theta}.

In the absence of 
residual systematic errors in galaxy shape measurement,
the expectation value of $\xi_{\pm}$ for two source populations is given, e.g. in 
\citet{Bartelmann2001}, by
\beq
\xi_{\pm,ab}(\theta) = \int_{0}^{\infty} \frac{\ell \rm d\ell}{2\pi} P_{\kappa, ab}(\ell) J_{0,4}(\ell \theta),
\label{eq:cosmic_shear_xis}
\eeq
where $J_{0, 4}(x)$ are the zero-th and fourth order Bessel functions of the first kind, respectively.
The zero-th order Bessel function should be adopted for $\xi_{+}$, while the fourth one is for $\xi_{-}$.

\citet{2002A&A...396....1S}
showed that the two point correlation functions of lensing shear are estimated 
in an unbiased way by averaging distortions over pairs of galaxies.
In practice, the estimator $\hat{\xi}_{\pm, ab}$ is calculated by
\begin{align}
\hat{\xi}_{\pm, ab}(\theta) =&
\frac{1}{(2{\cal R}^{(a)})(2{\cal R}^{(b)})} 
\frac{1}{N^{ab}_{p}(\theta)}\sum_{ij}w^{(a)}_{i}w^{(b)}_{j}
\Delta_{\theta}(\bd{\theta}_{i}-\bd{\theta}_{j})
\nonumber \\
&\times
\left(\epsilon^{(a)}_{t}(\bd{\theta}_{i})
\epsilon^{(b)}_{t}(\bd{\theta}_{j})
\pm
\epsilon^{(a)}_{\times}(\bd{\theta}_{i})
\epsilon^{(b)}_{\times}(\bd{\theta}_{j})\right), 
\label{eq:xi_pm_est_tomo} \\
N^{ab}_{p}(\theta) =& \sum_{ij}w^{(a)}_{i}w^{(b)}_{j}
\Delta_{\theta}(\bd{\theta}_{i}-\bd{\theta}_{j}),
\label{eq:eff_number_pairs_tomo}
\end{align}
where $w^{(a)}_{i}$ is weight related to shape measurement of $i$-th galaxy in catalogue ``$a$'',
${\cal R}^{(a)}$ is the responsivity\footnote{The factor of $2{\cal R}$ is needed for conversion of distortion to lensing shear $\gamma_{t}$, 
due to the definition of distortion used in this paper \citep[see Eq.~(19) in][]{2002AJ....123..583B}.} 
for catalogue ``$a$'',
$\Delta_{\theta}(\bd{\phi}) = 1$ for 
$\theta-\Delta \theta/2 \le \phi \le \theta-\Delta \theta/2$ 
and zero otherwise.
The expectation value of this estimator is evaluated by an ensemble average of the shear field $\gamma$ and it is known to be unbiased, 
i.e. $\langle \hat{\xi}_{\pm, ab}(\theta) \rangle = \xi_{\pm, ab}(\theta)$.

\subsection{Statistical uncertainties of 
cosmic shear correlation functions
}
\label{subsubsec:model_cov_cosmic_shear}

We then consider 
statistical uncertainties of cosmic shear correlation functions defined in
Section~\ref{subsec:cosmic_shear}.
For our observables, the covariance matrix is commonly adopted for 
an
evaluation of the statistical uncertainties.
The covariance matrix of two-point correlation function, denoted as
$\bd C$,
can be decomposed into three parts:
\beq
{\bd C} = {\bd C}_{\rm G} + {\bd C}_{\rm cNG} + {\bd C}_{\rm SSC}, \label{eq:cov_decomp}
\eeq
where ${\bd C}_{\rm G}$ represents the Gaussian 
error,
and 
other two terms denote the non-Gaussian errors arising from the four-point correlations of matter field in
large-scale structure.
The contribution ${\bd C}_{\rm cNG}$ arises from 
the four-point correlation within a given survey area \citep{2001ApJ...554...56C, 2009MNRAS.395.2065T}, 
while the term ${\bd C}_{\rm SSC}$ arises from correlations between
sub-survey (observable) modes 
and
super-survey (unobservable) modes comparable with or 
greater than the size of a survey window \citep[SSC stands for Super Sample Covariance;][]{2013PhRvD..87l3504T}.



\subsubsection{Gaussian covariance}

Following 
\citet{2008A&A...477...43J},
we use a formula for the Gaussian covariance expressed in terms of the convergence power spectrum:
\begin{align}
&{\rm Cov}_{\rm G}\left[\hat{\xi}_{X, ab}(\theta_{1}),  \hat{\xi}_{Y, cd}(\theta_{2})\right]
=
\frac{1}{\Omega_{s}}\int_{0}^{\infty}\frac{\ell {\rm d}\ell}{2\pi}\, J_{x}(\ell\theta_1)J_{y}(\ell\theta_2) \nonumber \\
&
\,\,\,\,\,\,\,\,\,\,\,\,\,\,\,\,\,\,
\,\,\,\,\,\,\,\,\,\,\,\,\,\,\,\,\,\,
\,\,\,\,\,\,
\times
\left(P^{\rm obs}_{\kappa, ac}(\ell)P^{\rm obs}_{\kappa, bd}(\ell)+P^{\rm obs}_{\kappa, ad}(\ell)P^{\rm obs}_{\kappa, bc}(\ell)\right), \label{eq:cov_G_xi_predict}
\end{align}
where $\Omega_s$ is the survey area and we use the notation of $J_{x}=J_0$ for $\xi_{X}=\xi_+$ and $J_{x}=J_4$ for $\xi_X=\xi_-$ and so on.
In the above, $P^{\rm obs}_{\kappa, ab}(\ell)$ represents the {\it observed} convergence power spectrum which includes
a contribution of intrinsic shape noise of source galaxies.
It is given by
\beq
P^{\rm obs}_{\kappa,ab}(\ell) = P_{\kappa, ab}(\ell) + \frac{1}{\bar{n}_{{\rm s}, a}}\left(\frac{\sigma_{\epsilon, a}}{2{\cal R}^{(a)}}\right)^2 \delta^K_{ab},
\label{eq:ps_obs}
\eeq
where $\delta^K_{ab}$ is the Kronecker delta function,
$\sigma_{\epsilon, a}$ and $\bar{n}_{{\rm s}, a}$ are the rms of intrinsic galaxy distortions per component and 
the mean source number density in catalogue ``$a$'', respectively.
Note $\sigma_\epsilon^2=\epsilon_{\rm rms}^2 + \sigma_e^2$ in terms of notations we introduce below: the shape noise arises from a sum of 
the ``intrinsic'' distortion and the measurement error.
On large scales comparable to the size of a survey window, the Gaussian prediction of Eq.~(\ref{eq:cov_G_xi_predict})
will be inaccurate because Eq.~(\ref{eq:cov_G_xi_predict}) ignores the boundary effect of survey window.
This finite area effect will be important for the HSCS16A data with sky coverage of $\sim100$ squared degrees.
\citet{Sato2011} developed the method to correct for the finite area effect of Gaussian covariance based on
the direct pair counting of observed galaxies. We extend their method to make it valid for tomographic analyses
in Appendix~\ref{apdx:finite_area_effect}.
A similar investigation of the impact of survey geometry on covariance estimation is found in \citet{2018MNRAS.479.4998T}.
It is worth noting that this study focuses on the case where the shape noise is less important, while \citet{2018MNRAS.479.4998T} mainly studied the effect of survey geometry on shape noise covariances.

\subsubsection{Non-Gaussian covariance}

The non-Gaussian covariance term of Eq.~(\ref{eq:xi_pm_est_tomo}) 
is
expressed as the sum of following two contributions:
\begin{align}
&{\rm Cov}_{\rm cNG}\left[\hat{\xi}_{X,ab}(\theta_{1}),  \hat{\xi}_{Y,cd}(\theta_{2})\right]
=\frac{1}{\Omega_s}\int\, \frac{{\rm d}^2\ell^{\prime}}{(2\pi)^2}\int\frac{{\rm d}^2\ell}{(2\pi)^2}
\nonumber \\
&
\,\,\,\,\,\,\,\,\,\,\,\,\,\,\,\,\,\,
\,\,\,\,\,\,\,\,\,\,\,\,\,\,\,\,\,\,
\,\,\,\,\,\,
\times 
J_{x}(\ell\theta_1)J_{y}(\ell^{\prime}\theta_2) T^{\rm cNG}_{\kappa, abcd}(\bd{\ell}, -\bd{\ell}, 
\bd{\ell}^{\prime}, -\bd{\ell}^{\prime}), \label{eq:cov_cNG_xi} \\
&{\rm Cov}_{\rm SSC}\left[\hat{\xi}_{X,ab}(\theta_{1}),  \hat{\xi}_{Y,cd}(\theta_{2})\right]
=\int_{0}^{\infty} \frac{\ell^{\prime}\, {\rm d}\ell^{\prime}}{2\pi}\int_{0}^{\infty}\frac{\ell \,{\rm d}\ell}{2\pi}
 \nonumber \\
&
\,\,\,\,\,\,\,\,\,\,\,\,\,\,\,\,\,\,
\,\,\,\,\,\,\,\,\,\,\,\,\,\,\,\,\,\,
\,\,\,\,\,\,
\times
J_{x}(\ell\theta_1)J_{y}(\ell^{\prime}\theta_2)
T^{\rm SSC}_{\kappa, abcd}(\ell, \ell^{\prime}, \Omega_s), \label{eq:cov_SSC_xi}
\end{align}
where $T_{\kappa}$ denotes the convergence trispectrum describing the four-point correlation function in Fourier space.
Using the Limber approximation, 
the connected trispectrum $T^{\rm cNG}_{\kappa, abcd}$ in Eq.~(\ref{eq:cov_cNG_xi}) can be computed
\citep[e.g.,][]{2001ApJ...554...56C, 2009MNRAS.395.2065T} as
\begin{align}
&T^{\rm cNG}_{\kappa, abcd}(\bd{\ell}_1, \bd{\ell}_2, \bd{\ell}_3, \bd{\ell}_4) = \int_{0}^{\chi_H}\!\!\frac{{\rm d}\chi}{r^6(\chi)}\, q_a(\chi) q_b(\chi) q_c(\chi) q_d(\chi)
\nonumber \\
&
\,\,\,\,\,\,\,\,\,\,\,\,\,\,\,\,\,\,
\,\,\,\,\,\,\,\,\,\,\,\,\,\,\,\,\,\,
\,\,\,\,\,\,
\times
T_{\rm m}\left(\frac{\ell_1}{r(\chi)}, \frac{\ell_2}{r(\chi)}, \frac{\ell_3}{r(\chi)}, \frac{\ell_4}{r(\chi)}, z(\chi)\right),
\end{align}
where $\bd{\ell}_1 + \bd{\ell}_2 + \bd{\ell}_3 + \bd{\ell}_4 = 0$
and $T_{\rm m}$ is the trispectrum of cosmic matter density field.
On the other hand, the SSC trispectrum $T^{\rm SSC}_{\kappa, abcd}$ can be expressed in \citet{2013PhRvD..87l3504T} as
\begin{align}
T^{\rm SSC}_{\kappa, abcd}(\ell_1, \ell_2, \Omega_s) 
=& \int_{0}^{\chi_H}\, \frac{{\rm d}\chi}{r^6(\chi)}\, q_a(\chi) q_b(\chi) q_c(\chi) q_d(\chi)
\nonumber \\
&\times
\frac{\partial P_{\rm m}(k_1, z)}{\partial \delta_{\rm bg}}\frac{\partial P_{\rm m}(k_2, z)}{\partial \delta_{\rm bg}} \sigma^2_W (z; \Omega_s), \label{eq:T_SSC}
\end{align}
where
\begin{align}
\sigma^2_W(z; \Omega_s) 
\equiv& \frac{1}{\Omega_s^2}\int\, \frac{{\rm d}^2\ell}{(2\pi)^2} P_{\rm m, L}(\ell/\chi(z), z)\, |\tilde{W}(\bd{\ell})|^2,
\label{eq:sigma_W}
\end{align}
and $k_{i} = \ell_{i}/\chi$,
$P_{\rm m, L}$ is the linear matter power spectrum,
and $\tilde{W}$ is the Fourier transform of survey window function.
Note we define the window function so that $\Omega_s = \int {\rm d}^2 \theta\, W(\bd{\theta})$.
In Eq.~(\ref{eq:T_SSC}), $\partial P_{\rm m}(k, z)/\partial \delta_{\rm bg}$ describes the response of power spectrum
to a fluctuation in background density $\delta_{\rm bg}$.
The SSC trispectrum arises from the four point correlation with squeezed quadrilaterals
including a shared infinite wavelength mode.

\section{Subaru Hyper Suprime-Cam Survey} 
\label{sec:data}

Hyper Suprime-Cam (HSC) is a wide-field imaging camera 
on the prime focus of the 8.2-meter Subaru telescope
\citep{2015ApJ...807...22M, 2018PASJ...70S...4A, 2018PASJ...70S...2K,2018PASJ...70S...3F, 2018PASJ...70S...1M}. 
Among three layers in the HSC survey,  
the Wide layer will cover 1400 ${\rm deg}^2$ 
in five broad photometric bands ($grizy$) 
over 5-6 years, with excellent image quality of sub-arcsec seeing.   
In this paper, we use a catalogue of galaxy shapes 
that has been generated for cosmological weak lensing analysis in the first year data release. 
The details of galaxy shape measurements and catalogue information are found in \citet{2018PASJ...70S..25M}.

In brief, the HSC S16A galaxy shape catalogue is based on the HSC Wide data taken from March 2014 to April 2016 with about 90 nights. 
We apply a number of cuts to construct a secure shape catalogue 
for weak lensing analysis \citep[see][for more details]{2018PASJ...70S..25M}. The selection criteria include
data selection with approximately full depth in all 
the 5 filters,
a conservative magnitude cut of $i<24.5$,
removal of galaxies with PSF modeling failures 
and those located in the disconnected regions. 
The sky around bright stars are masked \citep{2018PASJ...70S...7C}. 
As a result, the HSC S16A weak lensing shear catalogue covers 136.9~deg$^2$ that consists of 6 disjoint patches: XMM, GAMA09H, GAMA15H, HECTOMAP, VVDS, and WIDE12H. 
In the HSC S16A shape catalogue, the shapes of galaxies 
are estimated on the $i$-band coadded images using the 
re-Gaussianization PSF correction method \citep{2003MNRAS.343..459H}. 
This method has been applied to the Sloan Digital Sky Survey data, from which the systematics of the method are well understood
\citep{2005MNRAS.361.1287M,2013MNRAS.432.1544M}. 
In the method, the distortion of a galaxy image is defined as 
\begin{equation}
\bm{\epsilon}=(\epsilon_1,\epsilon_2)=\frac{1-(b/a)^2}{1+(b/a)^2}(\cos 2\phi, \sin 2\phi), \label{eq:def_epsilon}
\end{equation}
where $b/a$ is the minor-to-major axis ratio and $\phi$ is the
position angle of the major axis with respect to the equatorial
coordinate system.  
The shear of each galaxy, $\bm{\gamma}^{\rm (obs)}$, 
is estimated from the measured distortion $\bm{\epsilon}$ as follows:
\begin{equation}
\label{eq:obsshear}
\bm{\gamma}^{\rm (obs)}=\frac{1}{1+\langle m\rangle}\left(\frac{\bm{\epsilon}}{2{\cal
    R}}-\bm{c}\right),
\end{equation}
where ${\cal R}$ represents the response of our distortion definition
to a small shear \citep{2002AJ....123..583B} given by 
\begin{equation}
{\cal R}=1-\langle \epsilon_{\rm rms}^2 \rangle,
\end{equation}
where $\epsilon_{\rm rms}$ is the intrinsic root mean square (RMS)
distortion per component.
$\langle \cdot\cdot\cdot\rangle$ means the weighted
average with galaxy weight $w$ which is defined as the inverse
variance of the shape noise
\begin{equation}
\label{eq:weight}
w=(\sigma_{e}^2+\epsilon_{\rm rms}^2)^{-1},
\end{equation}
where $\sigma_{e}$ represents the shape measurement error for each
galaxy. The values $m$ and $\bm{c}$ represent the multiplicative
and additive biases in galaxy shapes. 
\citet{2018MNRAS.481.3170M} estimated both shape errors and biases
on object-by-object basis by using image simulations. We fully utilize the information of shape errors to construct mock catalogues of HSC S16A galaxy shapes (see Section~\ref{subsec:mock_shape} for details). In each patch, the survey windows is defined such that 1) the number of visits within {\tt HEALPix} pixels with {\tt NSIDE=1024} to be $(g,r,i,z,y)\geq(4,4,4,6,6)$ and the $i$-band limiting magnitude to be greater than 25.6, 2) the PSF modeling is good enough to meet our requirements on PSF model size residuals and residual shear correlation functions, 3) there are no disconnected {\tt HEALPix} pixels after the cut 1) and 2), and 4) the galaxies do not lie within the bright object masks. For details of defining these masks, see \citet{2018PASJ...70S..25M}.

\begin{figure}
\centering
\includegraphics[width=0.9\columnwidth]
{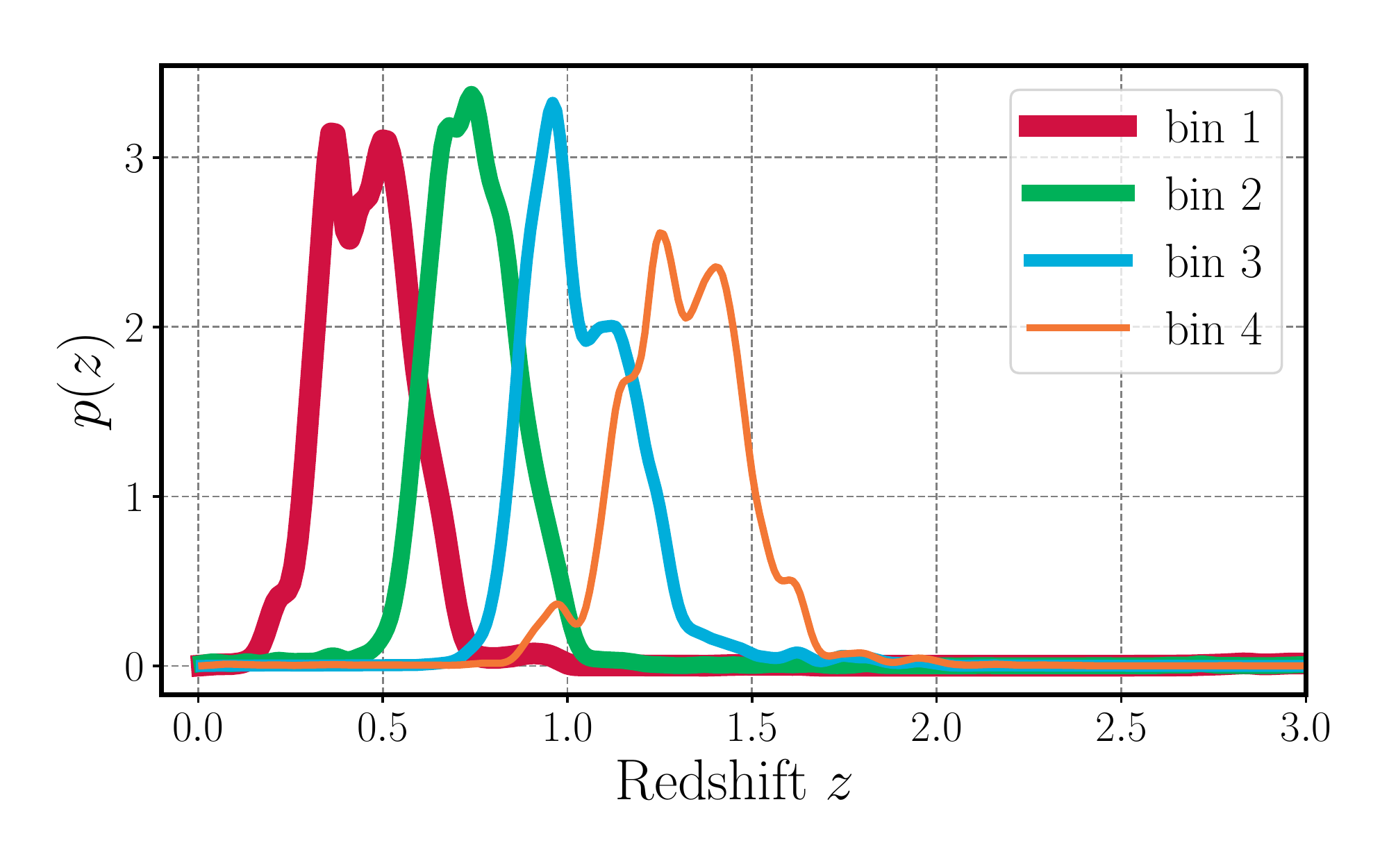}
\caption{
The stacked posterior redshift distributions of source galaxies in four tomographic redshift bins, 
which are used for the cosmic shear tomography analysis of HSC S16A data.
 Here we adopt the {\tt MLZ} catalogue of the photometric 
 redshifts in \citet{2018PASJ...70S...9T} as a default catalogue used in 
 the HSC mock catalogues.
}
\label{fig:HSCS16A_pz_4bins}
\end{figure}

The redshift distribution of source galaxies 
is estimated from the HSC five broadband photometry. 
\citet{2018PASJ...70S...9T} measured photometric redshifts (photo-$z$'s)
of galaxies in the HSC survey by using several different codes.
Among them, we choose the photo-$z$ 
with a machine-learning code based on self-organizing map ({\tt MLZ}) as a baseline\footnote{\ms{This is simply because 
the number of available photo-$z$ estimates is found to be largest in the photo-z catalogue based on {\tt MLZ}}.}.
To study the impact of photo-$z$ estimation with different methods, we consider three additional photo-$z$'s estimated 
from a classical template-fitting code ({\tt Mizuki}),
a neural network code using the CModel photometry ({\tt ephor}),
and a hybrid code combining machine learning with template fitting ({\tt frankenz}). When performing tomographic cosmic shear analyses, we divide the source galaxies into 4 subsamples 
by their {\it best} estimates \citep[see][]{2018PASJ...70S...9T}
of the photo-$z$'s ($z^{\rm best}$) in the redshift range from 0.3 to 1.5 as done in cosmic shear power spectrum analysis in the HSC S16A data \citep{2018arXiv180909148H}.
The redshift range of each tomographic bin is set to be
($0.3$, $0.6$), ($0.6$, $0.9$), ($0.9$, $1.2$), and ($1.2$, $1.5$) for the binning number from 1 to 4. 
Figure~\ref{fig:HSCS16A_pz_4bins} shows the stacked posterior distribution of photo-$z$ of source galaxies in four different tomographic bins.

\section{SIMULATIONS}\label{sec:sim}

\subsection{Full-sky simulation}

In order to construct the mock catalogues 
for weak lensing analyses in HSC, 
we utilize a large set of weak gravitational lensing simulations with all sky coverage. 
Here we briefly describe full-sky lensing catalogues\footnote{The full-sky light-cone simulation data are freely available for download at \url{http://cosmo.phys.hirosaki-u.ac.jp/takahasi/allsky_raytracing/}.}, 
while the details of these catalogues are found in \citet{2017ApJ...850...24T} 
\citep[also see][]{2017MNRAS.470.3476S}.
In \citet{2017ApJ...850...24T},
the authors performed a set of $N$-body simulations with $2048^3$ particles 
in cosmological volumes and used them to 
construct lensing and halo catalogues. 
They adopted the standard $\Lambda$CDM cosmology that is consistent with the WMAP cosmology
\citep{Hinshaw2013}.
The cosmological parameters are 
the CDM density parameter 
$\Omega_{\rm cdm}=0.233$, 
the baryon density 
$\Omega_{\rm b}=0.046$, 
the matter density 
$\Omega_{\rm m}=\Omega_{\rm cdm}+\Omega_{\rm b}
= 0.279$, 
the cosmological constant 
$\Omega_{\Lambda}=0.721$, 
the Hubble parameter
$h= 0.7$, 
the amplitude of density fluctuations
$\sigma_8= 0.82$,
and the spectral index
$n_s= 0.97$.
In the following,
we use 108 full-sky realizations in \citet{2017ApJ...850...24T}.

Full-sky weak gravitational lensing simulations
have been performed with the standard multiple lens-plane algorithm \citep[e.g.][]{2001MNRAS.327..169H, 2013MNRAS.435..115B, 2015MNRAS.453.3043S}. 
In this simulation, one can take into account the light-ray deflection on the celestial sphere
by using the projected matter density
field given in the format of spherical shell \citep[see, e.g.][for the similar approach]{2008MNRAS.391..435F}.
The simulations used the projected matter fields in 38 shells in total, each of which was computed by projecting 
$N$-body simulation realization over a radial width of $150\, h^{-1}{\rm Mpc}$, 
in order to make the light cone covering a cosmological volume up to $z=5.3$.
As a result, the lensing simulations
consist of shear field at 38 different source redshifts with angular resolution of 0.43 arcmin.
Each simulation data is given in the {\tt HEALPix} format \citep{2005ApJ...622..759G}.
The radial depth between nearest source redshifts is set to be $150\, h^{-1}{\rm Mpc}$ in comoving distance, corresponding to the redshift interval of $0.05-0.1$ for $z\simlt1$.

\subsection{Mock catalogues}
\label{subsec:mock_shape}
We here describe the details of creating the mock shape catalogues in HSC S16A from 
108 full-sky lensing simulations.
To do this, we follow the approach developed in \citet{2017MNRAS.470.3476S} \citep[also see][]{Shirasaki2014}.

In our mock catalogues, we incorporate the full-sky simulations with observed photometric redshift and angular position of real galaxies. 
Provided that real catalogue of source galaxies, where each galaxy contains information on the position (RA and dec), shape, redshift and the lensing weight, the procedures in production of mock catalogues consists of five steps as follows:

\begin{description}
 \item[(i)] Assign hypothetical RA and dec of survey window in the full-sky realization.
 \item[(ii)] Populate each source galaxy into one realization of the light-cone simulations according to its \ms{original} angular position and redshift. 
 \item[(iii)] Randomly rotate distortion of each source galaxy to erase the real lensing signal.  
 \item[(iv)] Simulate the lensing distortion effect on each source galaxy by adding the lensing contribution at each foreground lens plane
 \item[(v)] Repeat the steps (ii) -- (iv) for all the source galaxies
\end{description}

We then summarize some additional treatments to take into account the specific features in HSC on step-by-step basis. 

\begin{figure}
\centering
\includegraphics[width=0.80\columnwidth, bb=0 0 523 520]
{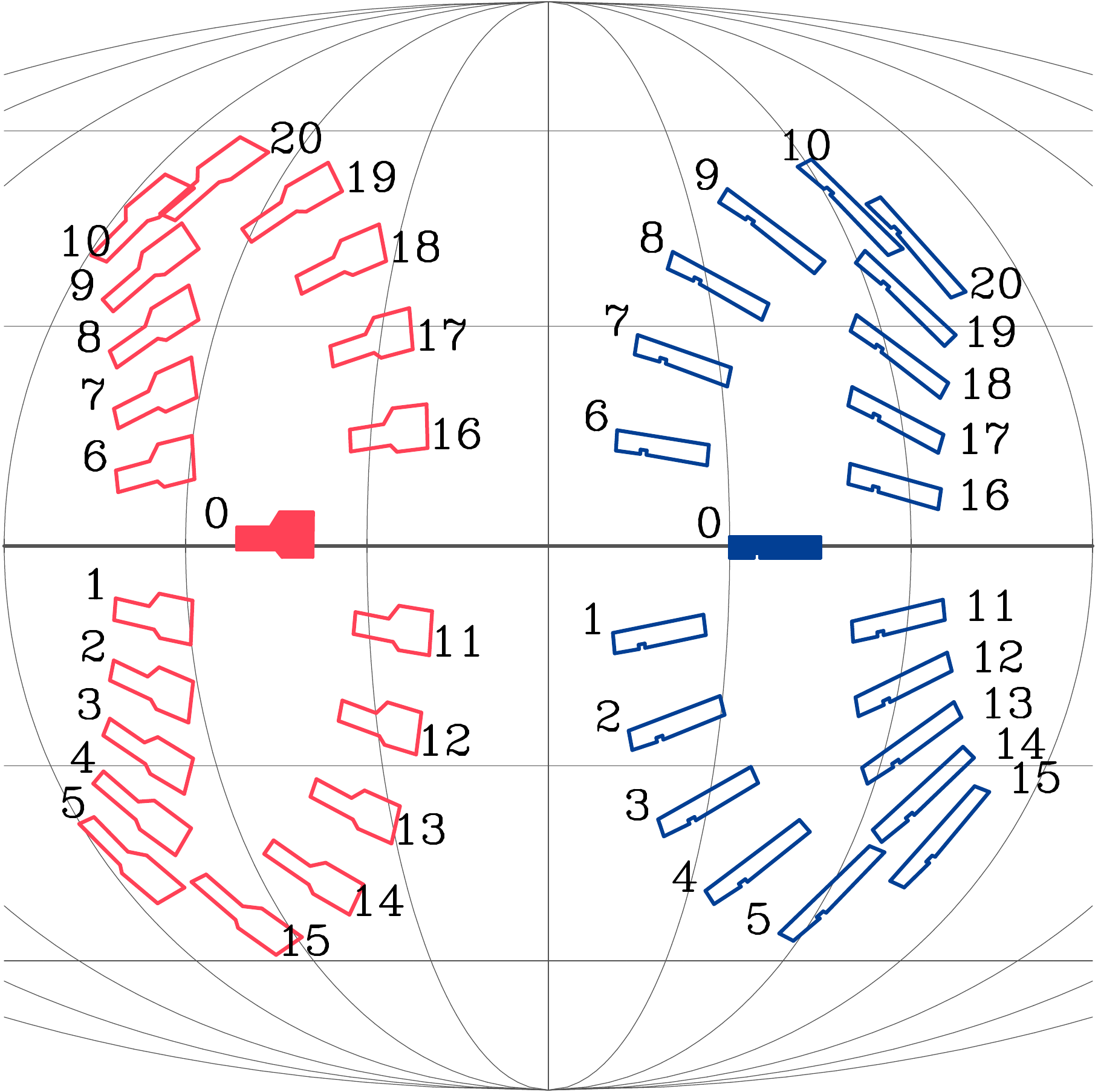}
\caption{
Distributions of the HSC fields in our full-sky mock catalogues. 
For each full-sky simulation, 
we take 21 realizations of the six distinct HSC fields. 
Here, for illustration purpose, we show only the two fields among six, the GAMA09 and GAMA15H fields, by red and 
blue regions. The filled regions show the locations without rotation, while the blank regions are their locations taken after 
rotations that are determined so that the different realizations are not overlapped with each other (see text for details). The indicies ranging from 0 to 20 represent each of the 21 realizations. Their relative locations of GAMA09 and GAMA15H in each 
realization are taken the same as those of the real HSC data.}
\label{fig:GAMA_rot}
\end{figure} 

\paragraph*{Step (i)} 
We pay a special attention 
to the positional relationship among HSC S16A regions. From a single full-sky simulation, we decide to make 21 mock shear catalogues by choosing the desired sky coverage of about 170 squared degrees 
which is the area when not taking into account masked regions and other cuts. 
Since 6 fields of HSC S16A are separated from each other, we define 21 different rotations on spherical coordinate so as to preserve the positional relationship among the HSC S16A fields.
Figure~\ref{fig:GAMA_rot} demonstrates the rotation of two HSC fields named as GAMA09H and GAMA15H. We set these rotations so that we do not use the same locations on full sky as possible. We found the area fraction of overlapped regions is about 2\% over 21 rotations. Note that we properly modify the simulated lensing field by taking into account the change of locally orthogonal coordinate system under a given rotation. 
Since 108 full-sky lensing simulations are available, we have $108 \times 21$ = 2268 realizations of each region of HSC S16A in total.
\ms{Under the rotation, we also keep the angular coordinates of individual source galaxies in the survey window. Hence, the source galaxies in our mock
have exactly same angular information as in the real catalogue.}

\paragraph*{Step (ii)} 
When 
injecting
each galaxy taken from the real
HSC source catalogue 
into
the light-cone simulation, 
we use the nearest pixel 
in the source plane at the nearest redshift, 
to those of the galaxy. 
In doing this
we 
use
a point
redshift estimate of 
each 
galaxy, taken from a random sampling of 
the posterior distribution of {\tt MLZ} photometric redshift estimate for the galaxy.
It is worth noting that we randomly generate 
redshift distributions
of source galaxies
when making a different realization of the mock catalogues.
Thus, our mock catalogues include effects of 
source galaxy properties 
(e.g., magnitudes, distortions and spatial  variations in the number densities), 
statistical uncertainties in photometric redshifts and the survey geometry. Although we use {\tt MLZ} photometric redshift estimate as our fiducial 
choice,
we also examine how the results are changed by using different photo-$z$ catalogues, 
{\tt ephor}, {\tt mizuki}, and {\tt frankenz}. We have created 2268 realizations 
for {\tt MLZ}, while we have generated
210 realizations 
for {\tt ephor}, {\tt mizuki}, and {\tt frankenz}.

\paragraph*{Step (iii)} 
Rotating the observed distortion of each galaxy
allows to eliminate 
real lensing distortions and \ms{intrinsic alignment signal} existing in galaxy images.
Assuming the amplitude in 
observed distortion is mainly determined 
by the intrinsic shape,
we can use this rotated distortion as a proxy of the intrinsic shape.
Nevertheless, the observed distortion 
contains an additional scatter due to the shape measurement error.
Following \citet{2018PASJ...70S..26O}
we simulate
an ``observed'' distortion of each galaxy
taking into account both
the intrinsic shape and measurement error.
We first rotate the distortion of individual galaxies $\bd{\epsilon}^{\rm obs}$
and obtain the rotated distortion as
$\bd{\epsilon}^{\rm ran}=\bd{\epsilon}^{\rm obs}e^{i\phi}$,
where $\phi$ is a random number between 0 and $2\pi$.
Here we need to distinguish the intrinsic distortion from the measurement error because 
the shear responsivity depends only on the intrinsic shape noise (rms).
We thus model the intrinsic shape $\bd{\epsilon}^{\rm int}$ and measurement error $\bd{\epsilon}^{\rm mea}$
from the following random generations:
\beq
\bd{\epsilon}^{\rm int} = 
\left(\frac{\epsilon_{\rm rms}}{\sqrt{\epsilon_{\rm rms}^2+\sigma_{\rm e}^2}}\right)\bd{\epsilon}^{\rm ran},
\,\,\,\,\,\,\,\,\,\,
\bd{\epsilon}^{\rm mea} = N_1 +i\, N_2,
\eeq
where 
$N_{i}$ is a random number drawn from a normal distribution with a standard deviation of $\sigma_{\rm e}$.
In the HSC shape catalogue, $\epsilon_{\rm rms}$ 
(parameter 
{\tt ishape\verb|_|hsm\verb|_|regauss\verb|_|derived\verb|_|rms\verb|_|e}) 
and $\sigma_{\rm e}$ 
(parameter
{\tt ishape\verb|_|hsm\verb|_|regauss\verb|_|derived\verb|_|sigma\verb|_|e}) are provided on object-by-object basis.

\paragraph*{Step (iv)}
We then 
obtain 
the mock ellipticy $\bd{\epsilon}^{\rm mock}$ 
\citep{1991ApJ...380....1M, 2002AJ....123..583B}:
\begin{align} 
\epsilon^{\rm mock}_{1} =& 
\frac{\epsilon^{\rm int}_{1}+
\delta_{1}+(\delta_{2}/\delta^2)
[1-(1-\delta^2)^{1/2}] (\delta_{1}\epsilon_{\rm int,
2}-\delta_{2}\epsilon^{\rm int}_{1})} {1+{\bm \delta}\cdot{\bm \epsilon}^{\rm int}} 
+\epsilon^{\rm mea}_{1}, \label{eq:emock1}\\ 
\epsilon^{\rm mock}_{2} =& 
\frac{\epsilon^{\rm int}_{2}+\delta_{2}+(\delta_{1}/\delta^2) [1-(1-\delta^2)^{1/2}]
(\delta_{2}\epsilon_{\rm int, 1}-\delta_{1}\epsilon^{\rm int}_{2})}
{1+{\bm \delta}\cdot{\bm \epsilon}^{\rm int}}
+\epsilon^{\rm mea}_{2}, \label{eq:emock2}
\end{align}
where ${\bm \delta}\equiv 2(1-\kappa){\bm
\gamma}/[(1-\kappa)^2+\gamma^2]$ and $\kappa$ and $\gamma$ are 
simulated lensing effects at the galaxy position,
taken from the light-cone simulation.
Note $\delta\simeq 2\gamma$ in the weak lensing regime
and we ignore any multiplicative and additive biases in mock catalogues in Eqs.~(\ref{eq:emock1}) and (\ref{eq:emock2}).
Nevertheless, one can include multiplicative and additive biases in our catalogues if needed,
since our mock catalogues share the same object ID with the real catalogue.
Except for Section~\ref{subsec:mbias}, we adopt Eqs.~(\ref{eq:emock1}) and (\ref{eq:emock2}) for simplicity.
We summarize the impact of non-zero multiplicative bias on the covariance of cosmic shear correlation function in Section~\ref{subsec:mbias}.

\section{RESULTS}\label{sec:res}

\subsection{Comparison with a theoretical model}
\label{subsec:comparion_model}

In this section, we compare the statistical property of clustering observables obtained from our mock catalogues
with its theoretical prediction in detail.

\subsubsection{Cosmic shear}

\begin{figure}
\centering
\includegraphics[width=1.0\columnwidth]
{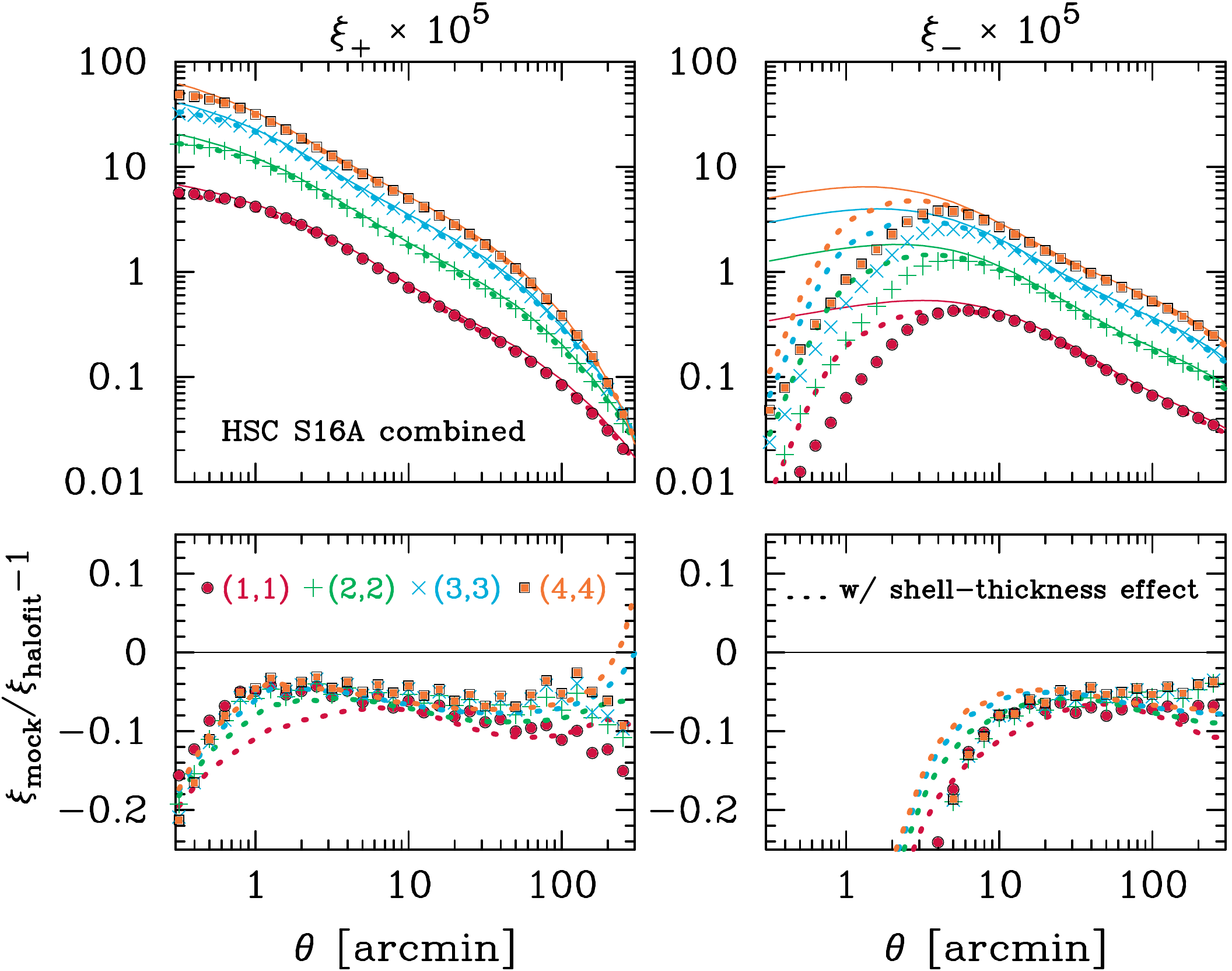}
\caption{Comparison of the analytical model predictions with those measured from 
the HSC mock catalogues for the cosmic shear two-point correlation functions in four 
tomographic redshift bins as shown in Fig.~\ref{fig:HSCS16A_pz_4bins}.
In the upper panels, the colored points show
the averages of 2268 mock realizations, while the solid lines are
the analytical predictions computed using the fitting formula of nonlinear matter power spectrum
\citep{Takahashi2012}. 
The fractional differences are shown in the lower panels.
The left and right plots are
the results for $\xi_{+}$
and $\xi_{-}$, respectively.
The dashed lines in each panel
show the theoretical predictions including effects of the finite thickness of lens planes in the line-of-sight integration 
that mimic the setup of ray-tracing simulation \citep{2017ApJ...850...24T} used in the mocks.
}
\label{fig:HSCS16A_shear_xis_mean}
\end{figure}

We validate the cosmic shear mock catalogues by comparing the cosmic shear 
correlation functions measured from the mocks with the theoretical expectation (Eq.~\ref{eq:cosmic_shear_xis}).
Figure~\ref{fig:HSCS16A_shear_xis_mean} shows 
the averaged $\xi_{\pm}$ 
from our 2268 realizations compared with the expectations. 
In this figure, we work with four different tomographic bins in source redshift selection as shown in Figure~\ref{fig:HSCS16A_pz_4bins}.
For simplicity, we consider the auto correlation functions of $\xi_{\pm}$ at single tomographic bin, i.e. $q_a = q_b$ in Eq.~(\ref{eq:P_kappa}).
Different colored points show the averaged $\xi_{\pm}$, while the lines represent 
the respective theoretical predictions that are computed using
the fitting formula of non-linear matter power spectrum in \citet{Takahashi2012}.
When measuring the correlation function in our mock catalogues, we use the public code {\tt Athena}\footnote{\url{http://www.cosmostat.org/software/athena}} \citep{2014ascl.soft02026K} and perform the logarithmic binning 
in the range of $0.281<\theta\, {\rm [arcmin]}<354$ with 31 bins.
We use the source galaxies in all of HSC S16A fields in Figure~\ref{fig:HSCS16A_shear_xis_mean}. 
When comparing the 
averaged $\xi_{\pm}$ with its expectation, we properly include
the selection bias $m_{\rm sel}$ due to cuts in the resolution factor 
and the responsivity correction due to the intrinsic distortion 
variations as a function of redshift in the theoretical model \citep[see Section~5.7][for details]{2018arXiv180909148H}.

First of all, the estimation of average $\xi$ in the mock catalogue should be reliable at $\theta\simgt1\, {\rm arcmin}$ for $\xi_+$ and $\theta\simgt10\, {\rm arcmin}$ for $\xi_-$ since the full-sky lensing simulations have an effective angular resolution corresponding to $\ell\simlt4000$ \citep{2017ApJ...850...24T}.
One can include this resolution effect in the model of lensing power spectrum (Eq.~\ref{eq:P_kappa}) as
\beq
P_{\kappa, ab}(\ell) \rightarrow P_{\kappa, ab}(\ell) \, \frac{1}{1+(\ell/\ell_{\rm sim})^2} \, \Theta(3 {\rm N_{SIDE}} - \ell), \label{eq:Pkappa_finite_angres}
\eeq
where $\Theta(x)$ is the Heaviside step function, 
$\ell_{\rm sim} = 1.6 \times {\rm N_{SIDE}}$ and ${\rm N_{SIDE}}=8192$ \citep{2017ApJ...850...24T}.
Apart from the angular resolution effect, 
we still find that the average $\xi_{\pm}$ can be different from its expectation value of Eq.~(\ref{eq:cosmic_shear_xis}). 
The differences of $\xi_{+}$ at $\theta\simgt1\, {\rm arcmin}$ 
and $\xi_{-}$ at $\theta\simgt10\, {\rm arcmin}$ 
in the bottom panels of Figure~\ref{fig:HSCS16A_shear_xis_mean} can be explained by 
the finite thickness effect of projected density shells in ray-tracing simulations \citep{2017ApJ...850...24T}.
We summarize the finite thickness effect on lensing power spectrum in Appendix~\ref{app:finite_thickness}.
Using Eqs.~(\ref{eq:Pkappa_finite_angres}), (\ref{eq:P_m_shell}), and (\ref{eq:Pkappa_sim_finitez}),
we obtain the theoretical model of $\xi_{\pm}$ including the effective angular resolution
and the finite thickness effect of projected density shells in the ray-tracing simulations. 
The dashed lines in Figure~\ref{fig:HSCS16A_shear_xis_mean} show the corrected version of $\xi_{\pm}$
and provide a better fit to the simulation results.

\begin{figure}
\centering
\includegraphics[width=1.0\columnwidth]
{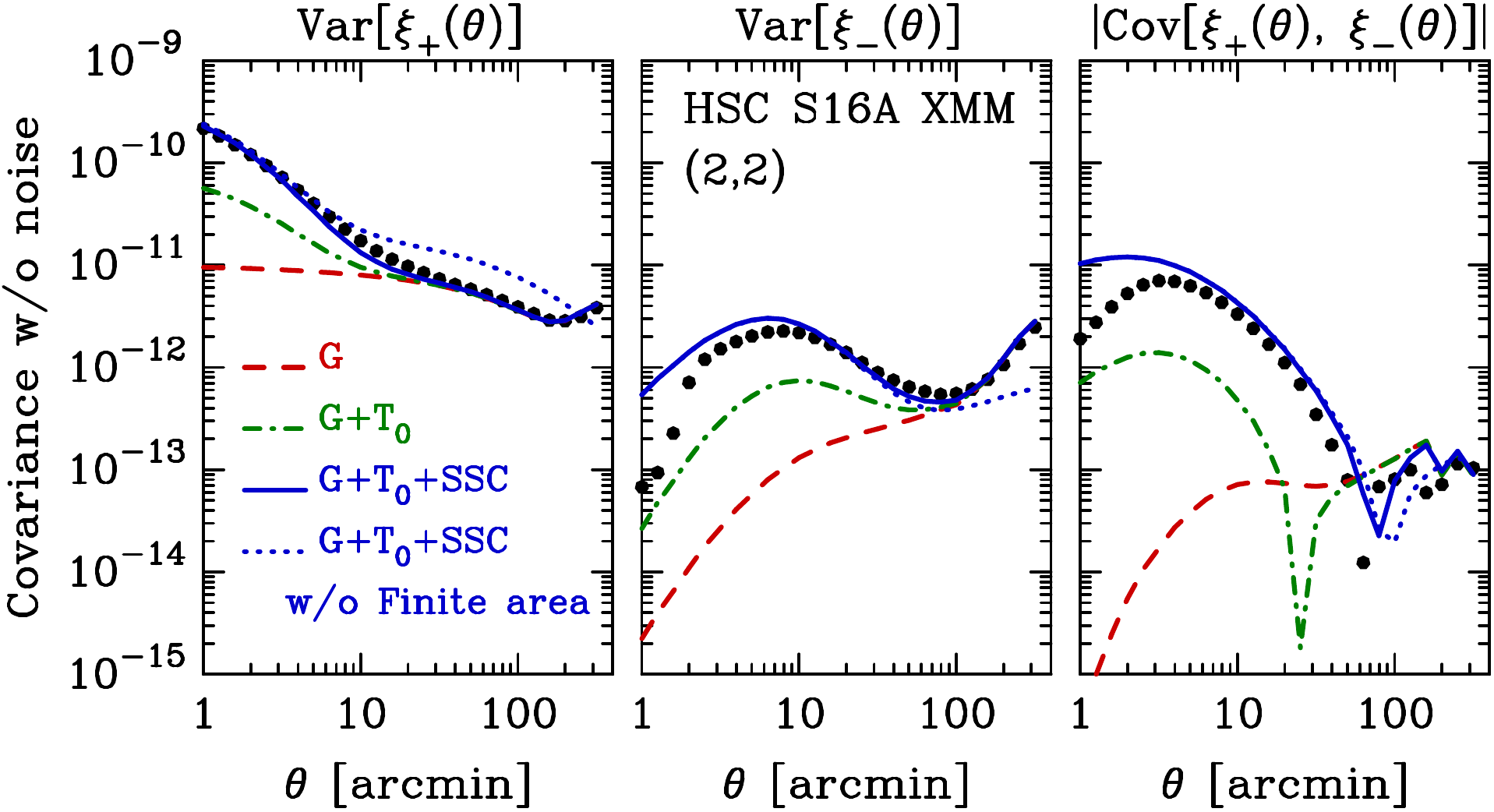}
\caption{
Comparison of the variance of cosmic shear correlation functions computed by the (semi-)analytical model and 
from the mock catalogues.
As an example we consider the correlation function for source galaxies in the second tomographic redshift bins for  
one of the HSC fields, the XMM field.
To study the impact of non-Gaussian sample variance, we ignore the shape noise contribution.
The left, middle, and right panels show
the variance of $\xi_+$,
the variance of $\xi_-$,
and the cross variance between $\xi_+$ and $\xi_-$,
respectively.
In each panel, the black point shows the mock results.
The colored lines show the different contributions in our analytical predictions; 
the red dashed lines show the Gaussian covariance, 
the blue solid lines are the non-Gaussian covariance, and the green dashed-dotted lines are the covariance without 
the SSC contribution. 
The dotted lines show the analytical predictions when ignoring effects of the survey boundary/footprint 
on the number of source galaxy pairs in the Gaussian variance calculation (see text for details and 
Appendix~\ref{apdx:finite_area_effect}). 
Note that the XMM field covers $\sim34.1\, {\rm deg}^2$.}
\label{fig:HSCS16A_XMM_shear_var_wonoise}
\end{figure}

We now discuss the covariance of cosmic shear correlation functions that resemble the HSC S16A data.
We first examine the theoretical model of the covariance matrix
in Section~\ref{subsubsec:model_cov_cosmic_shear}
by comparing the prediction with the covariance estimated from 2268 mock realizations.
To do this we use the mocks 
without galaxy shape noise, because we want to test whether the theoretical model can explain the mock covariance, 
from the linear to nonlinear scales.
\ms{For the mocks without shape noise, we use the lensing shear for a given source galaxy in the mocks and 
set unit weight when computing the two-point correlation, but keeping the angular and redshift information fixed as in the real catalogue.}
Figure~\ref{fig:HSCS16A_XMM_shear_var_wonoise}
shows the diagonal components of cosmic shear covariance
in one of the HSC S16A fields, the XMM field, in the absence of shape noises.
The black points in Figure~\ref{fig:HSCS16A_XMM_shear_var_wonoise}
show the mock covariance from 2268 realizations,
while different colored lines represent the theoretical prediction as in Section~\ref{subsubsec:model_cov_cosmic_shear}.
To compute the theoretical prediction,
we adopt
a halo-model approach \citep{2001ApJ...554...56C} as the relevant formulae are explicitly given in
Appendix~\ref{apdx:halo_model}.
For a computation of the Gaussian covariance, 
we properly take into account the effect of finite survey area or survey geometry. 
To be more precise, we properly estimate the number of pairs of two source galaxies, separated by a given angle,  
that can be taken from the HSC survey footprint (the XMM field considered here), as shown in Eq.~(\ref{eq:cov_pppp}).
We then estimate the weighted average of ensemble-averaged correlation functions, $\xi_\pm$, to obtain the prediction 
for the Gaussian covariance.
The blue solid lines are the analytical predictions which include
all the contributions, 
i.e. the Gaussian and non-Gaussian covariance contributions. 
The figure shows that the analytical predictions fairly well reproduce the mock results over the range of scales we consider.
On the other hand,
the blue dashed lines show the model predictions when ignoring the finite area effect in the Gaussian covariance calculation
or equivalently when using Eq.~(\ref{eq:cov_G_xi_predict}) (without shape noise), which is based on  a naive estimation of the number of source galaxy pairs. 
Comparison of the solid and dashed blue lines manifests that an accurate prediction of the Gaussian covariance requires to include the
effect of survey geometry in estimating the number of source galaxy pairs. 

\begin{figure}
\centering
\includegraphics[width=1.0\columnwidth]
{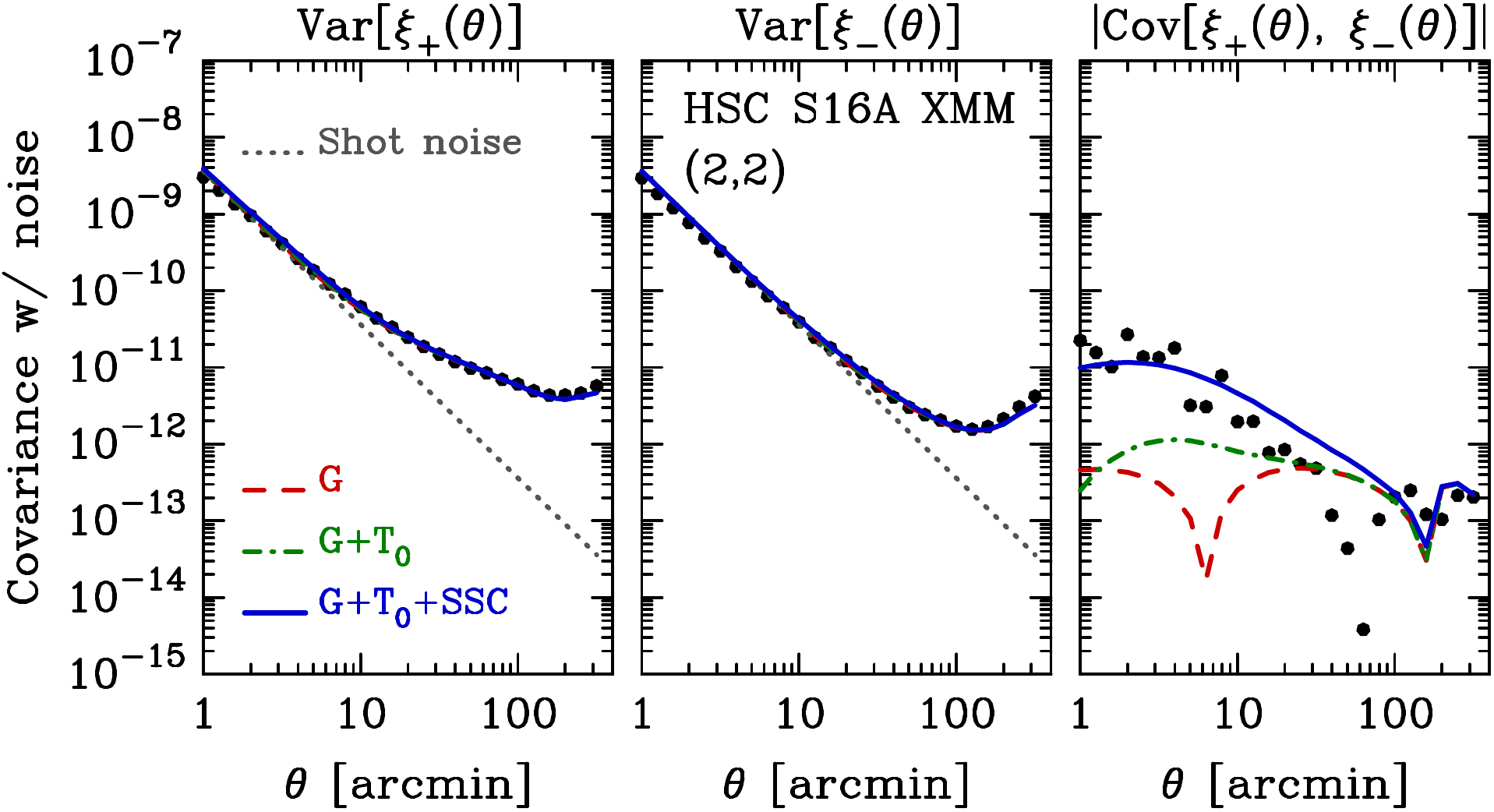}
\caption{
Similar to Figure~\ref{fig:HSCS16A_XMM_shear_var_wonoise}, but the intrinsic shape is included here.
The grey line represents the shot noise contribution alone. }
\label{fig:HSCS16A_XMM_shear_var_wnoise}
\end{figure}

Next, we consider the case including the shape noise contribution to the covariance. To compute 
the shape noise contribution in the analytical model, we use Eqs.~(\ref{eq:cov_G_xi_predict}) and (\ref{eq:ps_obs}) to compute 
the terms including $\sigma_\epsilon^4$ and $\sigma_\epsilon^2P_\kappa$, where we ignore the effect of survey geometry. 
To do this, we 
use the effective number density of source galaxies in the HSC S16A XMM field following Eq.~(1) in 
\citet{2012MNRAS.427..146H}, where we take into account the lensing weights for HSC galaxies.
For the sample variance of Gaussian covariance we use the same method in Figure~\ref{fig:HSCS16A_XMM_shear_var_wonoise} (i.e. the method given 
in Appendix~\ref{apdx:finite_area_effect}).
Figure~\ref{fig:HSCS16A_XMM_shear_var_wnoise} compares the analytical predictions with 
the mock results for the diagonal components in cosmic shear
covariance.
The simple analytical predictions fairly well reproduce the mock covariance which includes non-trivial observational effects 
such as the spatial and redshift distributions of source galaxies, the distribution of their distortions, and the lensing weights.
Compared to Figure~\ref{fig:HSCS16A_XMM_shear_var_wonoise},
the figure that the relative contribution of the non-Gaussian covariance to the shape noise covarinace is weaker
at $\theta<10\,$ arcmin. 
Nevertheless, it should be noted that the non-Gaussian covariance, especially the SSC term, 
is important for the off-diagonal elements as well as 
the cross covariance between $\xi_{+}$ and $\xi_{-}$, even in the presence of
the shape noise. 
We also note that the survey geometry effect on the sample varinace of Gaussian covariance needs to be properly taken into account.

\begin{figure}
\centering
\includegraphics[width=0.85\columnwidth]
{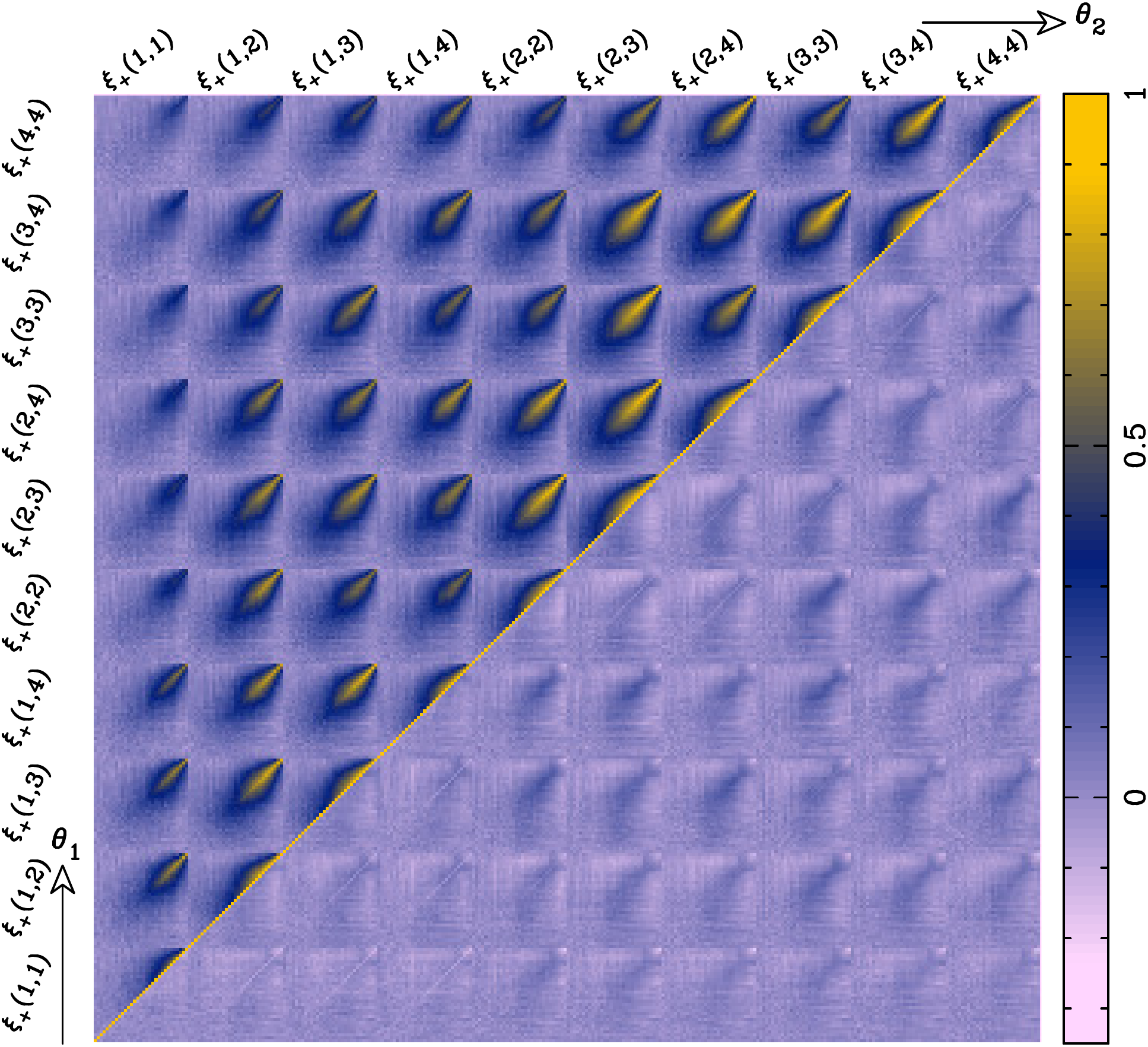}
\includegraphics[width=0.85\columnwidth]
{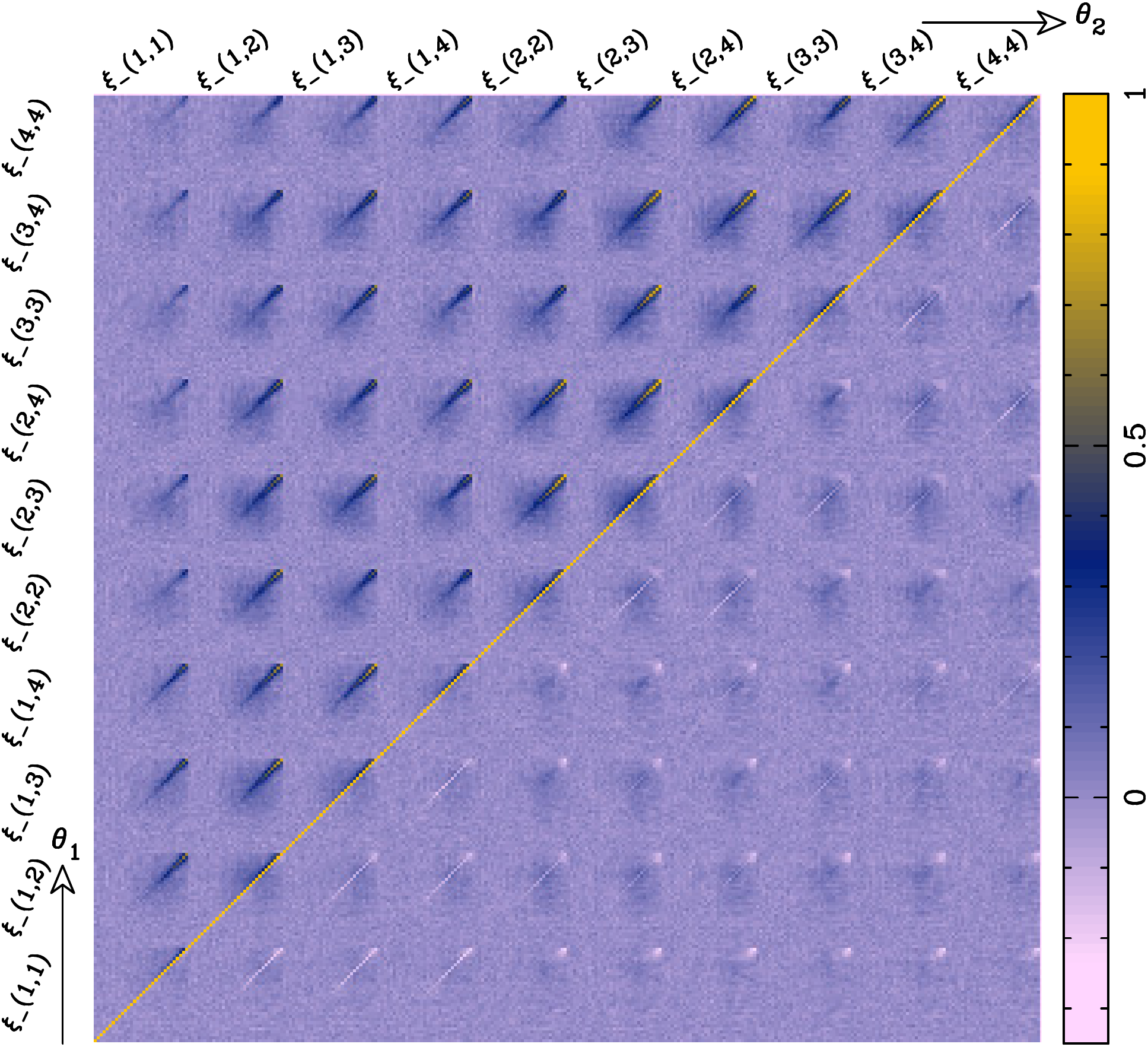}
\caption{
Comparison of the covariance matrices for $\xi_{\pm}$'s in four 
tomographic redshift bins (Fig.~\ref{fig:HSCS16A_pz_4bins}) computed by the 
analytical model and from the mock catalogues.
The top panel shows the result of $\xi_+$, while
the bottom is for $\xi_{-}$. 
We here consider 
the HSC S16A XMM field alone, and include the shape noise contribution. 
In each panel, the upper-left 
triangular elements show the mock covariance, while
\ms{the lower-right elements represent the difference between the mock result and our analytical prediction.}
Note the covariance matrix is normalized with the diagonal components.
\ms{Our analytic covariance is consistent with the mock counterpart with a $\sim20-30\%$ accuracy.}
}
\label{fig:HSCS16A_XMM_xis_cov}
\end{figure}

The upper or lower panel in
Figure~\ref{fig:HSCS16A_XMM_xis_cov} shows  the full covariance matrix 
of $\xi_+$ or $\xi_-$ including all the 4 tomographic bins, respectively.
In each panel, the upper-left triangle elements present
the mock covariance, 
while
the lower-right elements are 
\ms{the difference between the mock covariance and the analytical model prediction.}
For the Gaussian covariance computation in the analytical prediction,
we used the same method as 
in Figure~\ref{fig:HSCS16A_XMM_shear_var_wnoise}.
The analytical model covariance is again in a qualitatively nice agreement with the 
mock results including the off-diagonal components, in spite of the several approximations.
\ms{The accuracy of our model covariance is found to be the level of $20-30\%$.} 
Thus Figures~\ref{fig:HSCS16A_XMM_shear_var_wonoise} -- \ref{fig:HSCS16A_XMM_xis_cov} demonstrate that 
we have a good understanding of the nature of cosmic shear covariance, and we will below use the mock catalogues 
for a further discussion of the cosmic shear covariance.

\subsection{Field variations in clustering observables}

Since each of our mock catalogues, generated from full-sky realizations, properly consists of 
6 different fields in the HSC S16A data,
we here 
quantify the impact of field variations 
on the cosmic shear covariance\footnote{In this section, we focus on the field variation in the covariance.
In Appendix~\ref{app:est_fv}, we examine the field variation in the correlation functions themselves, 
$\xi_\pm$. We also note each field is separated from the others by $40-180$ deg.}.
To study this, 
we first measure the cosmic shear correlation function of source galaxies 
in a given combination of tomographic bins, denoted by
$a$ and $b$, from the field ``$f$'' of the $r$-th mock catalogue; 
we denote this correlation function as $\xi^{(r)}_{\pm, ab}(\theta;f)$.
We then combine all the correlation functions in the 6 fields to estimate
the ``full'' correlation function, denoted as $\xi_{\pm, ab}(\theta;{\rm comb})$, for the entire HSC S16A fields:
\beq
\xi^{(r)}_{\pm,ab}(\theta_i; {\rm comb})
= \frac{\sum_{f}N^{ab}_{p}(\theta_i; f)\, 
\xi^{(r)}_{\pm,ab}(\theta_i; f)}
{\sum_{f}N^{ab}_{p}(\theta_i; f)},
\eeq
where $N^{ab}_{p}(\theta_i; f)$ is 
the effective number of source galaxy pairs used in the cosmic shear correlation estimation, 
defined in Eq.~(\ref{eq:eff_number_pairs_tomo}).
Note that $N^{ab}_{p}(\theta_i; f)$ is the same
in all the
mock realizations since our mock catalogues use
the same spatial distribution of source galaxies as in the real catalogue.
The covariance matrix for the full correlation function can be 
estimated from all the mock realizations as
\begin{align}
&{\rm Cov}_{\rm full}
\left[\xi_{\pm,ab}(\theta_i), \xi_{\pm,cd}(\theta_j)\right]
= \frac{1}{N_{\rm rea}-1}\nonumber\\
&\,\,\,\,\,\,\,\,\,\,\,\,
\times\sum_{r=1}^{N_{\rm rea}}
\left\{\xi^{(r)}_{\pm,ab}(\theta_i; {\rm comb})-\bar{\xi}_{\pm,ab}(\theta_i; {\rm comb})\right\}\nonumber \\
&\,\,\,\,\,\,\,\,\,\,\,\,\,\,\,\,\,\,\,\,\,\,\,
\,\,\,\,\,\,\,\,\,\,\,\,\,\,\,\,\,\,\,\,\,\,\,
\times
\left\{\xi^{(r)}_{\pm,cd}(\theta_j; {\rm comb})-\bar{\xi}_{\pm,cd}(\theta_j; {\rm comb})\right\},
\label{eq:cov_full_HSC_cosmic_shear}
\end{align}
where
\begin{align}
&\bar{\xi}_{\pm,ab}(\theta_i; {\rm comb})
= \frac{1}{N_{\rm rea}}
\sum_{r=1}^{N_{\rm rea}} 
\xi^{(r)}_{\pm,ab}(\theta_i; {\rm comb}),
\end{align}
and $N_{\rm rea}=2268$ is the number of mock realizations.

For comparison, we also estimate the covariance matrix ignoring correlations between
the cosmic shear correlation functions in the different fields as
\begin{align}
&{\rm Cov}_{\rm no\, fv}
\left[\xi_{\pm,ab}(\theta_i), \xi_{\pm,cd}(\theta_j)\right] \nonumber \\
&= \frac{\sum_{f=1}^{N_{\rm field}} 
\, N^{ab}_{p}(\theta_i; f)\, N^{cd}_{p}(\theta_j; f)\, {\rm Cov}_{\rm each}\left[\xi_{\pm,ab}(\theta_i; f), \xi_{\pm,cd}(\theta_j;f)\right]}
{\left(\sum_{f=1}^{N_{\rm field}}N^{ab}_{p}(\theta_i; f)\right)
\left(\sum_{f=1}^{N_{\rm field}}N^{cd}_{p}(\theta_j; f)\right)},
\label{eq:cov_wofv_HSC_cosmic_shear}
\end{align}
where ${\rm Cov}_{\rm each}$ is the covariance of $\xi_{\pm}$
in each of the 6 HSC fields that is estimated from our mock catalogues in the field.

\begin{figure}
\centering
\includegraphics[width=0.80\columnwidth]
{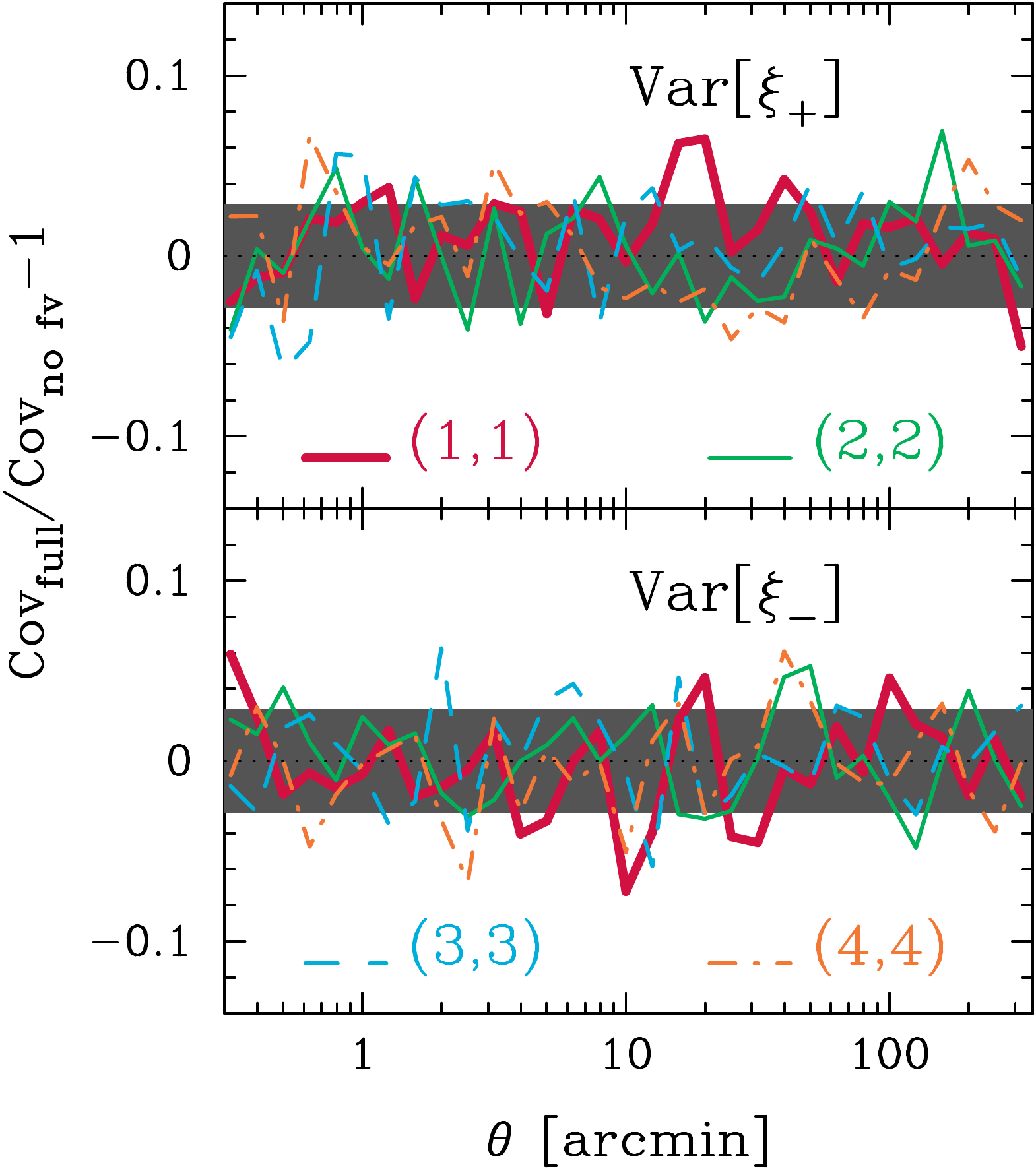}
\caption{
Impact of field variation on variance estimation of cosmic shear.
The upper panel shows the effect of field variation in the variance of $\xi_+$,
while the lower represents the case for $\xi_-$.
In each panel, we consider the fractional difference between the variance
with and without the field variation.
The variance including the field variation is defined in 
Eq.~(\ref{eq:cov_full_HSC_cosmic_shear}), while we estimate
the variance in the absence of field variations with Eq.~(\ref{eq:cov_wofv_HSC_cosmic_shear}).
Note that zero in y axis represents no impact of field variation.
Different colored lines in each panel present the results with
four different tomographic bins.
\ms{The thick solid, thin solid, dashed, dashed-dotted lines represent the first, second, third, and forth tomographic bin, respectively.}
\ms{Note that the grey filled region in each panel shows a statistical uncertainty in the variance estimation over 2268 realizations assuming a Gaussian distribution.}
}
\label{fig:HSCS16A_xis_var_fvimpact}
\end{figure}

Figure~\ref{fig:HSCS16A_xis_var_fvimpact} shows the fractional difference between the two covariances of Eqs.~(\ref{eq:cov_full_HSC_cosmic_shear})
and (\ref{eq:cov_wofv_HSC_cosmic_shear}).
To be specific, we plot the quantity of ${\rm Cov}_{\rm full}/{\rm Cov}_{\rm no\, fv}-1$.
Here we present only the diagonal components for the auto correlations of the same tomographic bins, $\xi_{\pm, aa}$.
It can be found that the effect of field variations on the covariance is not significant 
for the HSC 16A data. To be more precise, the fractional difference is at a level of $3-5\%$ over the range 
of angular scales.
Note that a statistical uncertainty in the variance estimation assuming a Gaussian distribution is
$\sqrt{2/N_{\rm rea}}\sim2.9\%$. 

\subsection{The impact of photo-$z$ errors on cosmic shear covariance}

Photo-$z$ errors are one of the most severe systematic effects on 
cosmic shear cosmology.
In this section, we study the impact of photo-$z$ errors
on the cosmic shear covariances using our mock catalogues.
Besides our fiducial set up, we also generate the mock catalogues of galaxy shapes using the posterior 
photo-$z$ distributions estimated using
three different photo-$z$
catalogues, {\tt ephor}, {\tt frankenz}, and {\tt mizuki}.
For this purpose we generate 210 realizations for each of the three catalogues.

\begin{figure}
\centering
\includegraphics[width=0.75\columnwidth]
{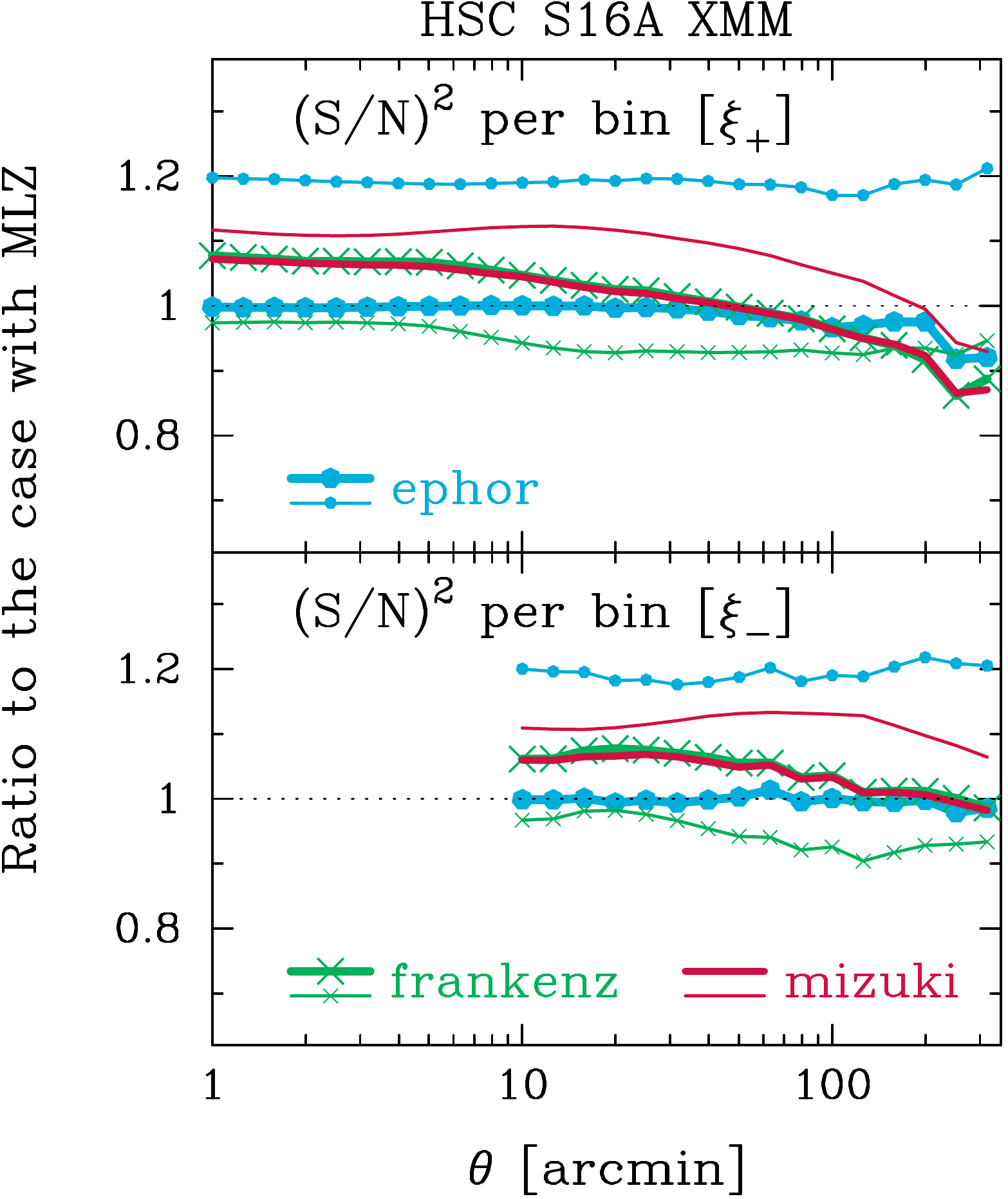}
\includegraphics[width=0.90\columnwidth]
{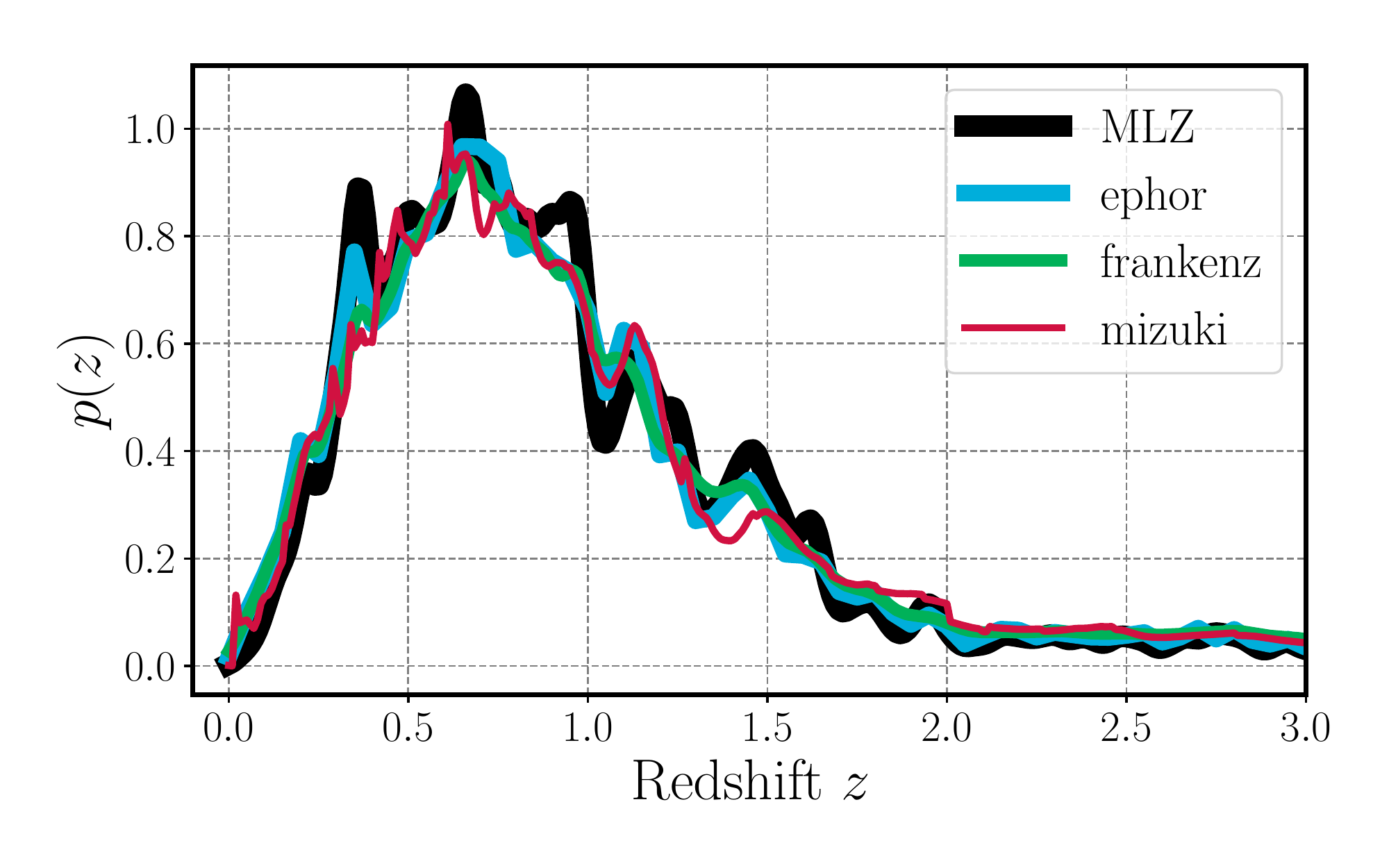}
\caption{
Impact of photo-$z$ posterior distributions on variance estimation of cosmic shear.
{\it Top}: The upper panel shows the signal-to-noise ratio ($S/N$) of $\xi_+$ for a given $\theta$,
while the lower represents the case for $\xi_-$.
In each panel, we consider the ratio of $(S/N)^2$
with three different photo-$z$ posterior distributions $p(z)$ to the counterpart with {\tt MLZ}.
\ms{Cyan lines with circle, green lines with cross symbol, and red solid lines represent the results 
with $p(z)$ from {\tt ephor}, {\tt frankenz}, and {\tt mizuki}, respectively.}
Hence, one in y axis means there are no differences in $S/N$ when one changes the $p(z)$ from {\tt MLZ}
to other.
Since we can not have a reasonable estimate of mean $\xi_-$ at $\theta\simlt10$ arcmin, 
we restrict the results for $\theta>10$ arcmin in the lower panel.
For comparison, the thin lines in upper and lower panels show the results when we change the variance according to photo-z methods 
but keep the signal fixed to the {\tt MLZ} case.
{\it Bottom}: The photo-$z$ posterior distributions from different methods used in the top panels.
}
\label{fig:HSCS16A_XMM_xis_var_diff_pz}
\end{figure}

To quantify the impact of photo-$z$ errors
on the cosmic shear covariance
we use the differential signal-to-noise ratio of
$\xi_{\pm}$ at each angular bin $\theta$, denoted as
$(S/N)^2_{\rm bin}$.
For this purpose we generate the mock catalogues based on the following method in order to 
make an apple-to-apple comparison of the results from different photo-$z$ catalogues. 
First, we use exactly the same number of source galaxies and the same realizations of 
full-sky simulations to simulate the lensing signals on each of source galaxies, for the 
four different photo-$z$ catalogues (the fiducial plus the three catalogues). To be more precise, 
we inject each of source galaxies, taken from the HSC S16A XMM field, into a given realization 
of the full-sky simulations. In doing this, we randomly assign a redshift of each source galaxy from 
the photo-$z$ posterior distribution for each of {\tt MLZ}, {\tt ephor}, {\tt frankenz}, and {\tt mizuki},
inject the galaxy into the nearest source plane of the simulation, and then simulate the lensing 
effect on the galaxy image (see Section~\ref{subsec:mock_shape}). We repeat this procedure for all the source galaxies.
We used 2,425,405 galaxies in total in each mock catalogue. 
The stacked posterior photo-$z$ distributions
for different methods
are shown in the lower plot in Figure~\ref{fig:HSCS16A_XMM_xis_var_diff_pz}.
The stacked redshift distributions display subtle different features from each other 
depending on which algorithm of photo-$z$ code to use.
Note that we do not consider a tomographic analysis in this subsection for simplicity. 

The upper plot
in Figure~\ref{fig:HSCS16A_XMM_xis_var_diff_pz}
shows the ratio of 
$(S/N)^2_{\rm bin}$, obtained from each of the mock catalogues
based on the different photo-$z$ codes, relative to
that of {\tt MLZ} photo-$z$ catalogue.
The upper panel represents the results for $\xi_{+}$,
while the lower is for $\xi_{-}$.
In each panel, the thick line shows the ratio of $(S/N)^2_{\rm bin}$ when we include the changes of $\xi_{\pm}$ and their variance
by different $p(z)$, while the thin line corresponds to the case changing the variance alone and using the same signals as the {\tt MLZ} case.
It can be found that the different photo-$z$ catalogues yield slightly different $(S/N)^2_{\rm bin}$
by a $\sim 5$-$10\%$ level.

From comparison of the thick and thin lines in the upper plot, 
it can be found
that the change in the variance due to the different photo-$z$ methods, keeping the signal fixed to the same, yields 
a 10-20\% change in $(S/N)^{2}_{\rm bin}$. If further including the change in the signal due to the different photo-$z$ catalogues, the two 
effects are somewhat compensated, and the net change in $(S/N)^{2}_{\rm bin}$ becomes smaller (within a 10\% level over the range of scales).
It would be worth noting that the shape noises in the sample used in Figure~\ref{fig:HSCS16A_XMM_xis_var_diff_pz}
are much smaller than one in tomographic analyses as in Figure~\ref{fig:HSCS16A_XMM_shear_var_wnoise}.
In Figure~\ref{fig:HSCS16A_XMM_xis_var_diff_pz}, the sample variance (including non-Gaussian terms) dominates 
the variance of $\xi_{\pm}$ over the angular range of $\sim3-100$ arcmin.
Hence, when changing $p(z)$, the signal and their variance change in a similar way and the $(S/N)^2_{\rm bin}$ is less affected effectively.
Assuming less source number density as in Figure~\ref{fig:HSCS16A_XMM_shear_var_wnoise}, 
we expect that the $(S/N)^2_{\rm bin}$ at $\theta \simlt 10$ arcmin can change with a level of $10-20\%$ according to different $p(z)$
since the shape noise dominates the variance at those angular scales.
Nevertheless, the $(S/N)^2_{\rm bin}$ at the sample-variance dominated regime will be still less affected as shown in Figure~\ref{fig:HSCS16A_XMM_xis_var_diff_pz} as long as one use an appropriate $p(z)$ for both of signals and their covariance.

\subsection{Signal-to-noise ratio}

A large set of the HSC mock catalogues enables us to 
estimate an expected, cumulative signal-to-noise ratio for a measurement of the 
cosmic shear correlation functions from the HSC S16A data.
Assuming four tomographic redshifts bins (see Figure~\ref{fig:HSCS16A_pz_4bins}),
we construct a data vector of $\bd{D}$
from different combinations of the cosmic shear correlation functions 
given as a function of angular bins and tomographic bins,
$\xi_{\pm, ab}(\theta_{i})$ 
over
the angular range of $10\le \theta_{i}\, [\rm arcmin]\le 100$.
We adopt up to 11 angular bins 
in the angular range and have 
10 correlation functions for each of $\xi_{\pm}$
at each angular bin that are available from combinations of 
the 4 redshift bins; therefore, 
the dimension of data vector 
$220=10\times 11\times 2$ at maximum.
The cumulative signal-to-noise ratio up to a maximum angular scale $\theta_{\rm max}$
is defined as
\beq
\left(\frac{S}{N}\right)^2_{\rm cum} \left[\theta_{\rm max}\right]
= \sum_{
\theta_{\rm min}\le\{\theta_i,\theta_j\}\le\theta_{\rm max}}
\sum_{p,q} \,^{t}\bar{{\bd D}}(\theta_i; p) 
\, {\bd C}^{-1} \, \bar{{\bd D}}(\theta_j; q),
\label{eq:S2N_cumulative_cosmic_shear_tomo}
\eeq
where $\theta_{\rm min}=10\, \rm arcmin$,
${\bd C}$ is the covariance matrix of data vector ${\bd D}$,
$\bar{{\bd D}}$ is the expectation value of ${\bd D}$, 
\ms{$^t{\bd D}$ is the transposed vector of ${\bd D}$,}
and
$p$ and $q$ represent the indexes of  tomographic bins in $\xi_{\pm}$.
For the data vector $\bar{{\bd D}}$,
we use the averaged 
$\xi_{\pm}$ over 2268 mock realizations.
We also study the differential signal-to-noise ratio at a given angular bin, $\theta$, defined as
\beq
\left(\frac{S}{N}\right)^2_{\rm bin}
= \sum_{p,q} \,^{t}\bar{{\bd D}}(\theta; p) 
\, {\bd C}^{-1} \,\bar{{\bd D}}(\theta; q).
\label{eq:S2N_bin_cosmic_shear_tomo}
\eeq
Note that,  
when
estimating the inverse matrix
${\bd C}^{-1}$ in Eqs.~(\ref{eq:S2N_cumulative_cosmic_shear_tomo})
and (\ref{eq:S2N_bin_cosmic_shear_tomo}), 
we included the correction factor of
$N_{\rm rea}/[N_{\rm rea}-N_D-1]$
proposed in \citet{Hartlap2007},
where $N_{\rm rea}$ is the number of the realizations and $N_D$ 
is the dimension of data. 

\begin{figure}
\centering
\includegraphics[width=0.80\columnwidth]
{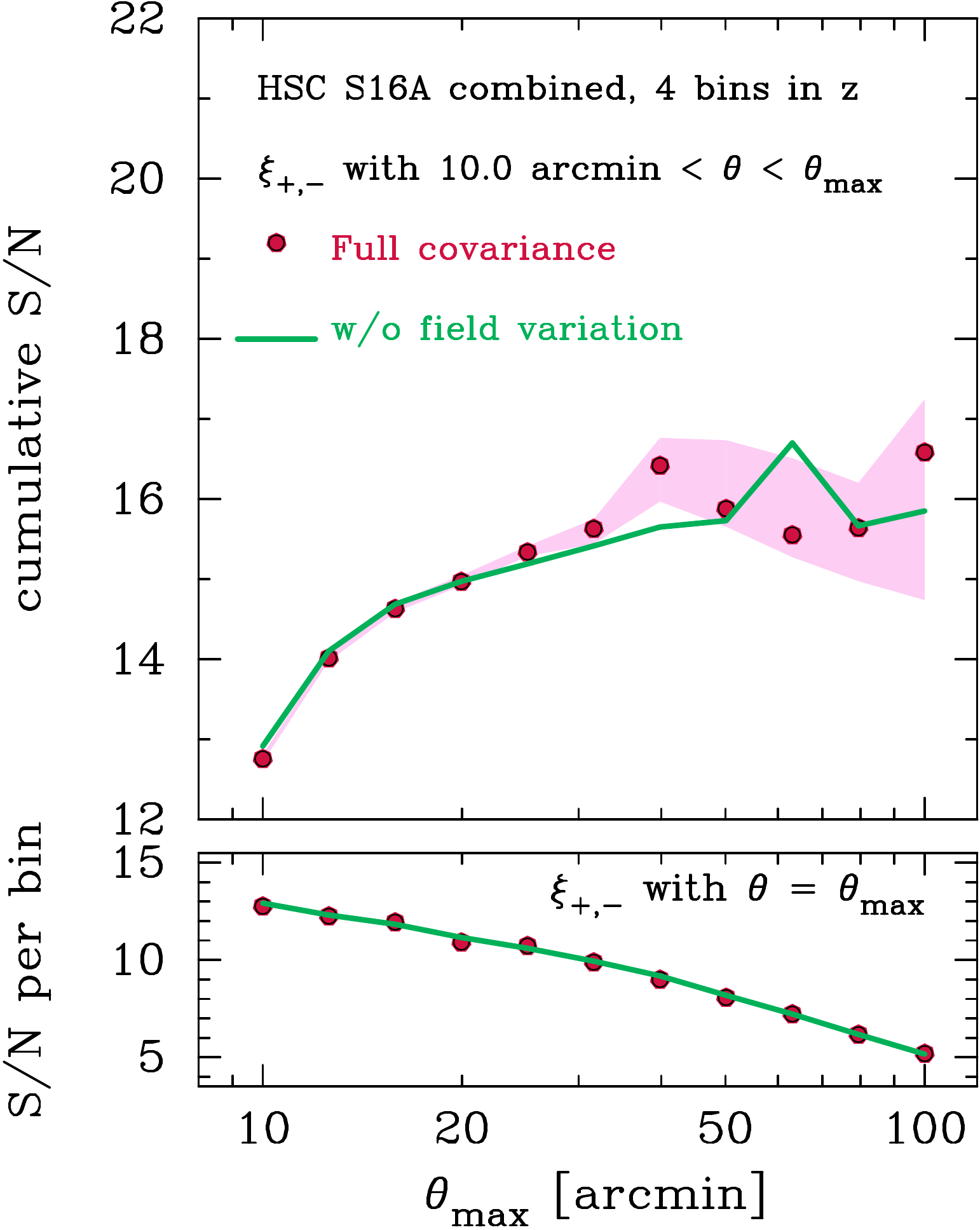}
\caption{
The expected signal-to-noise ratio of tomographic cosmic shear analysis with 4 bins in HSC S16A.
The upper panel shows the cumulative signal-to-noise ratio ($S/N$) as a function of maximum angular scales $\theta_{\rm max}$.
In the upper panel, we set the minimum scale to be 10 arcmin and use 10 tomographic $\xi_+$ and $\xi_-$.
Pink filled region in the upper panel show the results from 108 bootstrap samplings of 2247 measurements of $\xi_{\pm}$.
The lower panel represent the $S/N$ for a given angular scale, but we put together with 10 possible tomographic bins.
In each panel, the red points show the $S/N$ with full covariance from 2268 mock realizations,
while the green lines present the $S/N$ in the absence of field variations.
}
\label{fig:HSCS16A_xis_s2n}
\end{figure}

Figure~\ref{fig:HSCS16A_xis_s2n} shows the results for the cumulative or differential 
$S/N$ value expected for the HSC S16A data.
In this figure, we use two different covariance matrices; one is the full covariance evaluated with Eq.~(\ref{eq:cov_full_HSC_cosmic_shear}),
and another is the covariance in the absence of field variation among six different HSC S16A fields, defined as Eq.~(\ref{eq:cov_wofv_HSC_cosmic_shear}).
The figure shows the cumulative signal-to-noise ratio in HSC S16A cosmic shear analyses
is expected to be $16-17$ if the cosmology assumed in our simulations is correct.
The field variation is found to be less important for estimation of 
signal-to-noise ratio at a given angular scale, 
but it can induce a $6-7\%$ difference 
in the total signal-to-noise ratio 
if the correlation functions $\xi_{\pm}$ at scales around 1 degree are included.
To study the scatter in the signal-to-noise ratio, we perform bootstrap sampling of 2247 mock catalogues 108 times\footnote{\ms{Note that we extract 21 HSC S16A fields from a single full sky and the number of full-sky simulations is 108. Hence, it is easiest to construct 108 bootstrap samples of $107\times21=2247$ mocks in our configuration.}}
and then measure the cumulative signal-to-noise ratio for each bootstrap realization.
The pink filled region represents the bootstrap results and indicate the scatter in the evaluation of the cumulative signal-to-noise ratio in our mocks.
\ms{The difference between the red point and the green line in the upper panel of Figure~\ref{fig:HSCS16A_xis_s2n}
is still comparable to the bootstrap scatter, indicating that the difference can be consistent with a statistical fluke.}
We further comment on the result of cumulative $S/N$ in Figure~\ref{fig:HSCS16A_xis_s2n}.
We find the field variation can change the off-diagonal components in the covariance of $\xi_{+}$
and the cross covariance between $\xi_{+}$ and $\xi_{-}$. 
Including the field variance is found to  increase the cumulative $S/N$ of $\xi_{+}$ in our tomographic analysis 
with a level of $2-3\%$ compared to the case in the absence of field variation, but it does not affect $\xi_{-}$.
We also confirm that the cumulative $S/N$ of $\xi_{+}$ or $\xi_{-}$ monotonically increases as a function of $\theta_{\rm max}$,
while the cross covariance between $\xi_{+}$ and $\xi_{-}$ would induce the complex feature in the cumulative $S/N$ of four-tomographic $\xi_{\pm}$ at $\theta_{\rm max}>40-50$ arcmin.
In addition, the bootstrap scatter in the cumulative $S/N$ of $\xi_{\pm}$ 
at $\theta_{\rm max} > 50$ arcmin is of an order of $5-10\%$, while the Gaussian uncertainty of covariance with 2247 realizations is $\sim2.9\%$.
This implies that the degree-scale $\xi_{\pm}$ in tomographic analysis of the HSC S16A data will be non-Gaussian 
and may break a common approximation in cosmological likelihood analyses (also see the following subsection).

\subsection{Likelihood function for parameter inference} 

\begin{figure}
\centering
\includegraphics[width=0.85\columnwidth]
{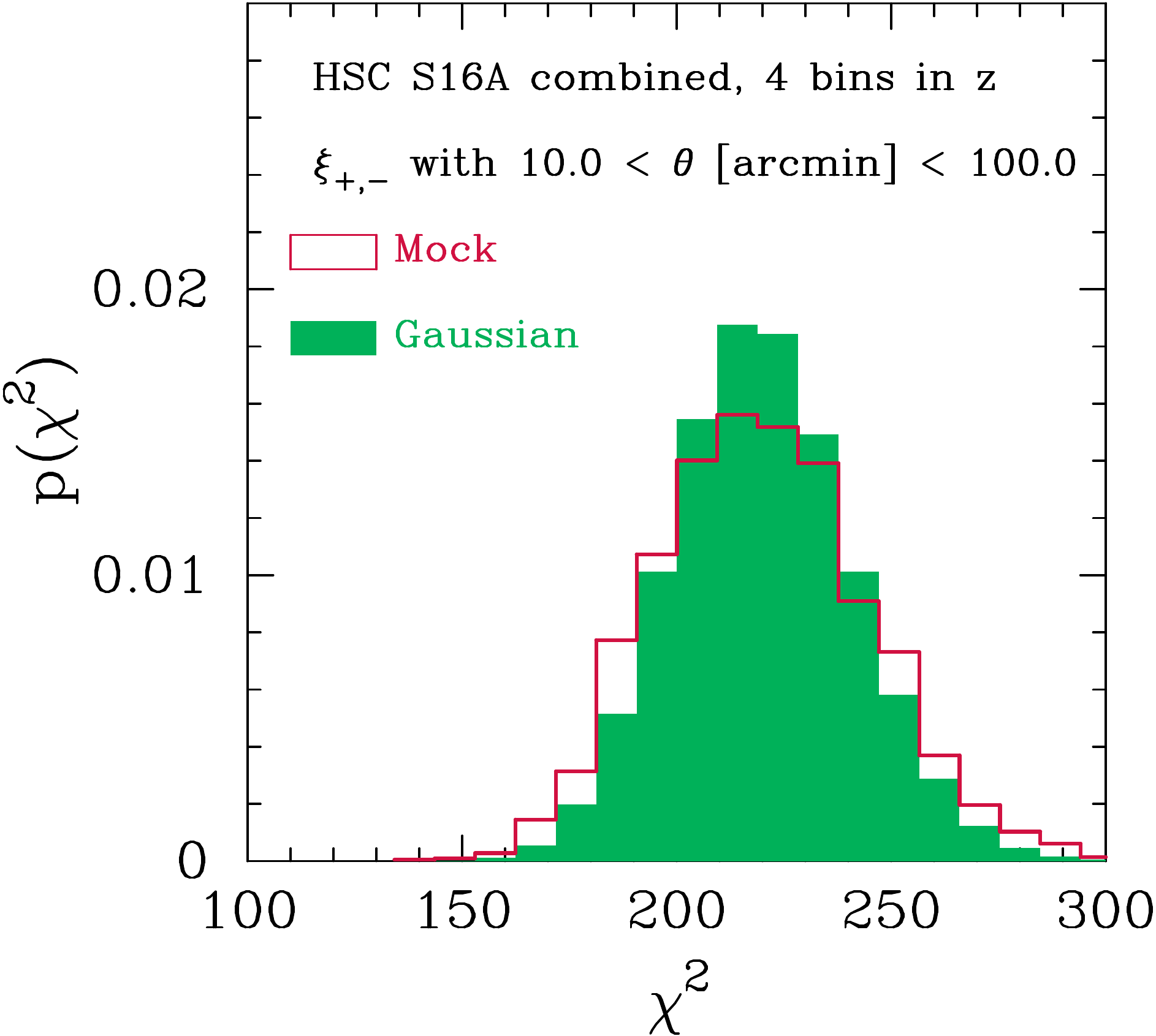}
\caption{
Probability distribution function (pdf) of chi squared value in HSC cosmic shear tomography.
The definition of chi squared is found in the text.
We here assume four different tomographic bins and combine 6 patches in HSC S16A.
Red open histogram shows the pdf obtained from 2268 mock realizations,
while green filled histogram represents the expectation from Gaussian cosmic shear correlations.
In this figure, we use the correlation functions $\xi_{\pm}$ in the range of $10\le\theta\, [{\rm arcmin}]\le100$.
We have two correlations of $\xi_{\pm}$ 
with 11 bins in $\theta$ and 10 bins in redshifts, leading the degrees of freedom to be $2\times11\times10=220$.
}
\label{fig:HSCS16A_xis_chi_sq_dist}
\end{figure}

\begin{figure}
\centering
\includegraphics[width=0.85\columnwidth]
{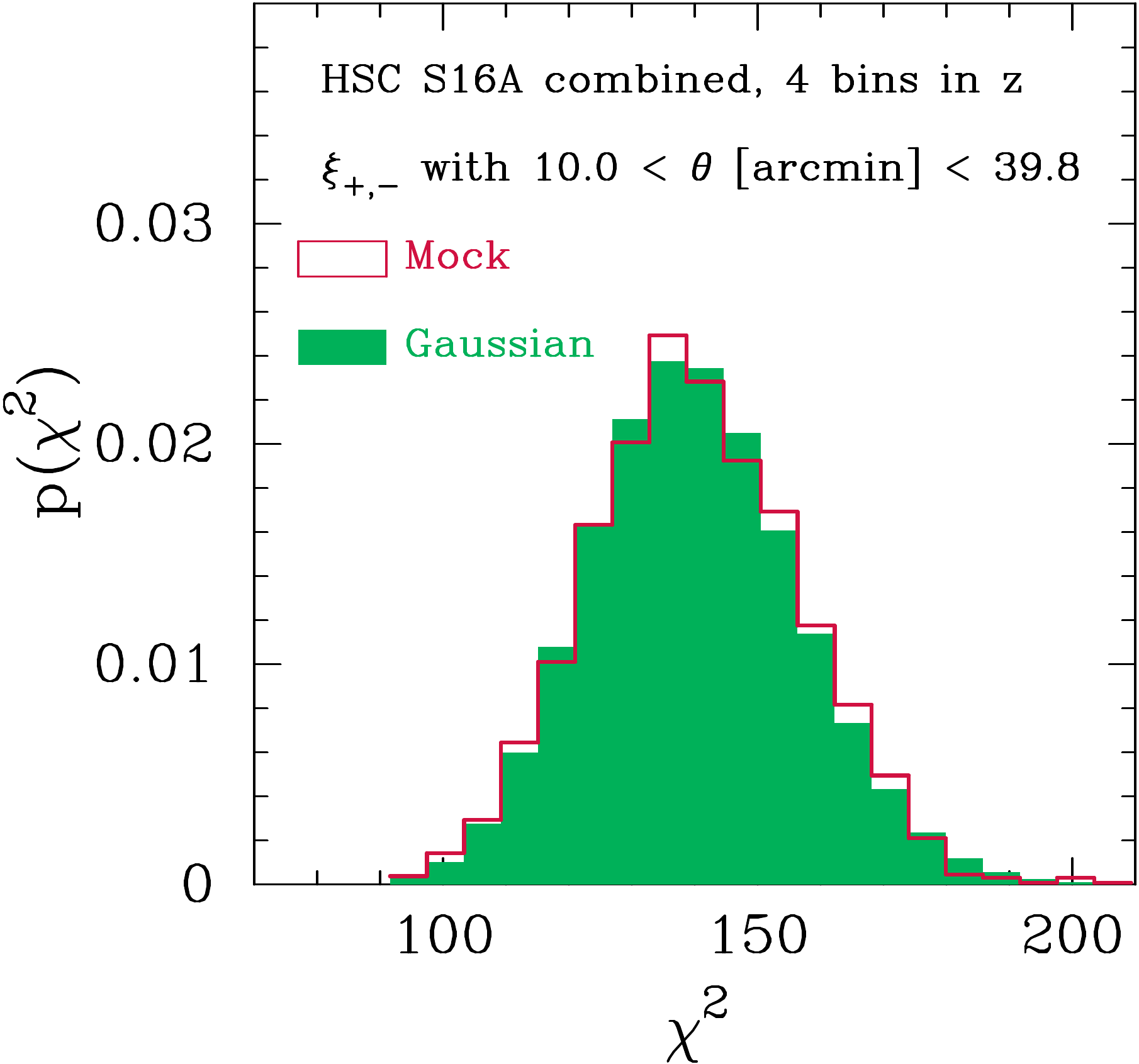}
\caption{
Similar to Figure~\ref{fig:HSCS16A_xis_chi_sq_dist}, but we restrict the angular range of $10-40$ arcmin.
Here we have two correlations of $\xi_{\pm}$ 
with 7 bins in $\theta$ and 10 bins in redshifts, leading the degrees of freedom to be $2\times7\times10=140$.
}
\label{fig:HSCS16A_xis_chi_sq_dist_10to40}
\end{figure}

The cosmological parameter estimation with cosmic shear
requires a likelihood function of correlation functions $\xi_{\pm}$.
Although the likelihood function is assumed to be Gaussian in practice,
recent studies claim that the likelihood function of cosmic shear correlation functions can be skewed \citep[e.g.][]{2018MNRAS.477.4879S} and Gaussian assumption
may affect the cosmological parameter estimation with cosmic shear
\citep[e.g.][]{2009A&A...504..689H,2010PhRvL.105y1301S}.
Hence, it is worth studying if the non-Gaussian likelihood function
can matter in the cosmic shear analyses with HSC S16A. 
To quantify the non-Gaussianity in cosmic-shear likelihood and evaluate its impact on parameter estimation, 
we define so-called chi squared quantity $\chi^2$ as
\beq
\chi^{2} \equiv \sum \, 
^{t}\left({\bd D}-\bar{{\bd D}}\right)\, 
{\bd C}^{-1}\, \left({\bd D}-\bar{{\bd D}}\right), \label{eq:chisq}
\eeq
where the summation runs over all the angular bins in the range of $10 \le\theta[{\rm arcmin}]\le 100$ and possible tomographic bins
over four different source redshift bins.
Note that the quantity of $\chi^2$ can be set for individual realizations of mock catalogue and it should follow the $\chi^2$ distribution with 220 degrees of freedom
if ${\bd D}$ follows Gaussian.

Figure~\ref{fig:HSCS16A_xis_chi_sq_dist} shows the histogram of $\chi^2$ 
that is computed from 2268 mock catalogues for a hypothetical measurement of the HSC S16A cosmic shear 
correlation functions.
In this figure, green filled histogram represents the Gaussian prediction,
while red line shows the mock results. 
This figure shows the non-Gaussian likelihood of cosmic shear correlations
could broaden a confidence level in a parameter estimation (e.g. an amplitude parameter such as $S_{8}$).
We find that the mean value of $\chi^2$ over 2268 realizations is 219.03
and it is in good agreement with simple Gaussian expectation, while
the variance of $\chi^2$ in our mock catalogues 
is found to be 627.86, corresponding to 
1.42 times as large as Gaussian prediction. 
For a 95\% confidence level, the interval in mock $\chi^2$
should range from 181.173 to 230.495, while their Gaussian counterparts
are 209.686 and 228.273, respectively.

The red open histogram in Figure~\ref{fig:HSCS16A_xis_chi_sq_dist} is also found to be explained by
a modified $\chi^2$ distribution as
\beq
p(\chi^2, N_{\rm dof}) = f(\chi^2 - N_{\rm dof} + k, k), \label{eq:mod_chisq_dist}
\eeq
where $f(x, k)$ is the $\chi^2$ distribution with $k$ degrees of freedom.
When setting to $N_{\rm dof}=220$, 
we find $k=315$ can explain the broadening of the histogram of $\chi^2$ in Figure~\ref{fig:HSCS16A_xis_chi_sq_dist}.
According to the above, we will discuss the effective number of degrees of freedom in the HSC S16A cosmic shear analyses.
Since the signal-to-noise ratio of $\chi^2$ with $k$ degrees of freedom is given by $k/\sqrt{2k}$, 
the effective degrees of freedom $k_{\rm eff} \sim 220^2/315 = 153.6$ gives the similar signal-to-noise ratio of $\chi^2$ in our mock analyses.
This computation of effective degrees of freedom implies that the effective number of angular bins per 
tomographic bin will be $153.6/2/10 \sim 7.68$.
Since we have 11 bins in $\theta$ in Figure~\ref{fig:HSCS16A_xis_chi_sq_dist}, 
the largest $3-4$ bins in $\theta$ may be effectively less important for the evaluation of $\chi^2$.
A principal component analysis will be required to study how 
the different $\xi_{\pm}$ are correlated with each other and 
how many independent bins contribute to most of the information contents in more details \citep[e.g.][]{2013MNRAS.429..344K}.
Note that we do not include the correction factor to 
invert $\bd C$ as in \citet{Hartlap2007} in Figure~\ref{fig:HSCS16A_xis_chi_sq_dist}. 
When we include the correction, the red open histogram shifts to right with a level of $10\%$, but the 95\% confidence level is still wider than
its Gaussian counterpart. \ms{It is also worth noting that Eq.~(\ref{eq:mod_chisq_dist}) is not a unique expression to characterize the non-Gaussianity in likelihood function of cosmic shear two-point correlations. Appropriate treatment of non-Gaussian likelihood may be needed to extract all information of cosmic shear two-point correlations, while it is beyond the scope of this paper.}




On the other hand, Figure~\ref{fig:HSCS16A_xis_chi_sq_dist_10to40} shows the histogram of $\chi^2$ 
in 2268 realizations of mock cosmic shear analyses when we use the information at the angular range of $10-40$ arcmin.
Comparing with Figures~\ref{fig:HSCS16A_xis_chi_sq_dist} and \ref{fig:HSCS16A_xis_chi_sq_dist_10to40},
we conclude the degree-scale $\xi_{\pm}$ can broaden the width of the histogram in $\chi^{2}$ 
in our mock catalogues.
Once we focus on the angular range of $\theta=10-40$ arcmin, the histogram of $\chi^2$ in our mocks 
follows the expected $\chi^2$ distribution \citep[Similar results are found in][]{2018MNRAS.477.4879S}.
Figures~\ref{fig:HSCS16A_xis_chi_sq_dist} and \ref{fig:HSCS16A_xis_chi_sq_dist_10to40} demonstrate
that our mock catalogues allow to set the angular scales at which a Gaussian likelihood approximation is valid.

\subsection{Impact of non-zero multiplicative bias}\label{subsec:mbias}

So far, we assumed the multiplicative bias in the shear of each object in mock catalogues to be zero.
In this subsection, we examine the impact of non-zero multiplicative bias in the tomographic correlation analysis of cosmic shear
in the HSC S16A. Note that we still assume zero additive biases in this subsection.

Since the correction of multiplicative bias is valid for the average shape over a given sample of source galaxies,
we need to be careful when including the multiplicative bias on object-by-object basis.
In this paper, we propose the following modification in lensing shear when incorporating with simulation and observed data sets
as in Eqs.~(\ref{eq:emock1}) and (\ref{eq:emock2}):
\begin{align}
{\bm \gamma} \rightarrow (1+\langle m \rangle)\,{\bm \gamma},
\end{align}
where ${\bm \gamma}$ represents the lensing shear from full-sky ray-tracing simulation
and $\langle m \rangle$ is the average multiplicative bias for the galaxy sample of interest.
When working on four tomographic bins, we will have four different values of $\langle m \rangle$.
For galaxies at $i$-th tomographic bin, we use the corresponding multiplicative bias $\langle m_i \rangle$
to produce the mock distortion. This simple procedure enables us to keep the shear responsivity fixed regardless 
of the value of multiplicative bias.

At the lowest-order level of Eqs.~(\ref{eq:emock1}) and (\ref{eq:emock2}), the mock distortion can be expressed as
\begin{align}
{\bm \epsilon}^{\rm mock} \simeq {\bm \epsilon}^{\rm int} + {\bm \epsilon}^{\rm mea} + 2{\cal R}(1+\langle m \rangle){\bm \gamma}.
\end{align}
To obtain an unbiased estimate of lensing shear from ${\bm \epsilon}^{\rm mock}$, we will use ${\bm \epsilon}^{\rm mock}/(2{\cal R})/(1+\langle m \rangle)$, but this correction leads the effective shape noise $({\bm \epsilon}^{\rm int} + {\bm \epsilon}^{\rm mea})/(2{\cal R})$ changes
by a factor of $(1+\langle m \rangle)^{-1}$.
Therefore, including non-zero multiplicative biases remains the estimator of $\xi_{\pm}$ unbiased as long as one include the correction of $(1+\langle m \rangle)^{-1}$ to the observed distortion, while the covariance of $\xi_{\pm}$ should be affected by non-zero $\langle m \rangle$.

\begin{figure}
\centering
\includegraphics[width=1.0\columnwidth]
{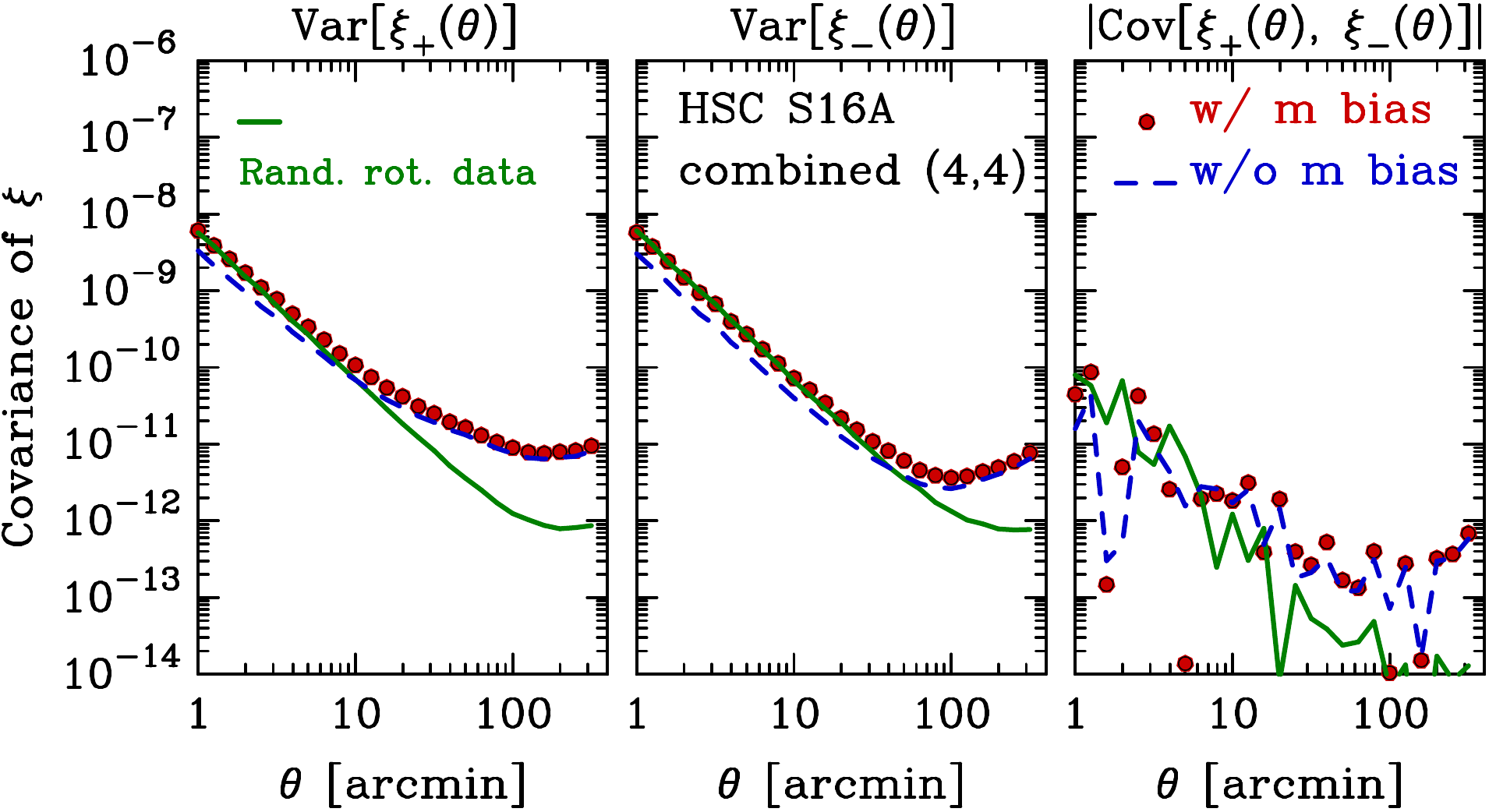}
\caption{
Impact of multiplicative bias on the covariance of cosmic shear correlation functions.
As a representative example, we consider the correlation function for source galaxies in the fourth tomographic redshift bin.
In each panel, the red points show the diagonal covariance estimated from the mock catalogues with non-zero multiplicative bias,
while the blue dashed line represent the mock covariance when the multiplicative bias is set to be zero in mock catalogues.
As a reference, the green solid line in each panel shows the covariance by the actual data but each shape in the catalogue
is randomly rotated. Hence, the green solid line should represent the shape noise term in covariance alone.
Including multiplicative bias can boost the amplitude of shape noise in cosmic shear analysis,
but the large-scale covariance is less affected by the presence of non-zero multiplicative bias.
}
\label{fig:HSCS16A_xis_var_mbias}
\end{figure}

Figure~\ref{fig:HSCS16A_xis_var_mbias} highlights the impact of non-zero multiplicative bias on the covariance of $\xi_{\pm}$ in the HSC S16A.
In this figure, we show the variance of $\xi_{\pm}$ and the cross variance between two when including non-zero multiplicative biases
for the fourth tomographic redshift bin of source galaxy sample in the HSC S16A.
The red points in the figure show the results in the presence of non-zero $\langle m \rangle$,
while the blue dashed line corresponds to the cases with $\langle m \rangle = 0$.
Note that we find $\langle m \rangle$ has a negative value over six HSC S16A fields and four tomographic bins.
Hence, the effective shape noise should increase in our four-tomographic analysis when we include the non-zero $\langle m \rangle$.
As a reference, the green dashed lines in this figure show the covariance estimated from randomly rotated shapes in real HSC S16A data.
Comparing with the red points and the green solid lines, we confirm that the small-scale covariance can increase due to the change 
of the effective shape noise. On the other hand, the large-scale covariance is less affected by the presence of non-zero $\langle m \rangle$,
since the sample variance should dominate at degree scales and it is independent of the amplitude of shape noise.

\begin{figure}
\centering
\includegraphics[width=0.80\columnwidth]
{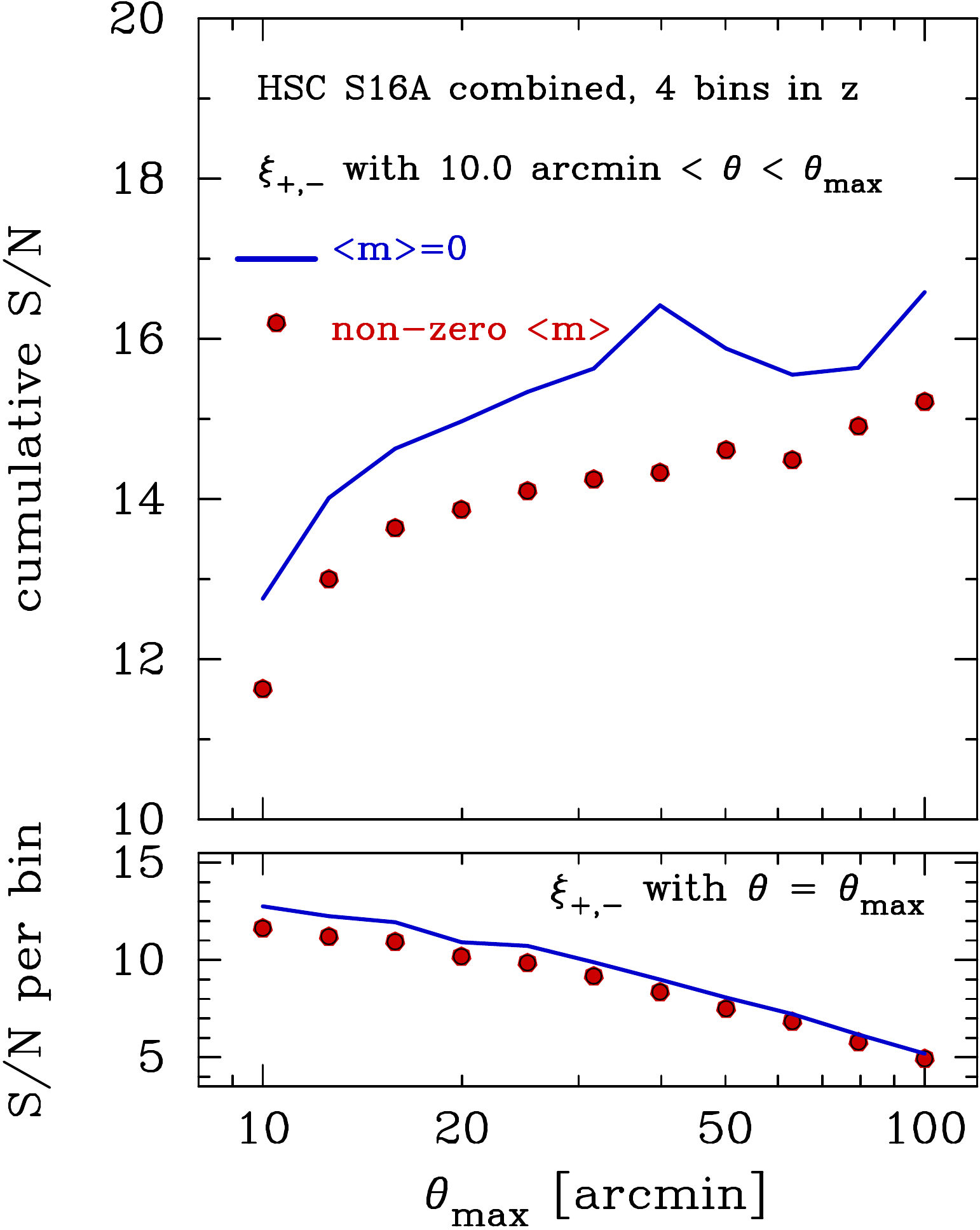}
\caption{
Similar to Figure~\ref{fig:HSCS16A_xis_s2n}, but we include the non-zero multiplicative biases in our HSC mock catalogues.
In each panel, the red points show the results in the presence of non-zero multiplicative biases
and the blue solid lines are for zero multiplicative biases. For the tomographic cosmic shear analysis in HSC S16A,
including the non-zero multiplicative bias can degrade the cumulative signal-to-noise ratio with $\simeq20\%$ level.
}
\label{fig:HSCS16A_xis_s2n_mbias}
\end{figure}

Figure~\ref{fig:HSCS16A_xis_s2n_mbias} demonstrates the impact of non-zero multiplicative bias of signal-to-noise ratio of cosmic shear
analysis in the HSC S16A, which is an indicator of cosmological information contents in the analysis.
When combining all HSC S16A fields and four tomographic bins in the angular range of $10-100$ arcmin, we find the non-zero multiplicative bias
can affect the cumulative signal-to-noise ratio by $\sim20\%$ in the HSC S16A.
It would be worth noting that the impact of multiplicative bias on the shape noise depends on the sign of multiplicative bias.
If the multiplicative bias can have a positive value, the effective shape noise will be reduced by a factor of $(1+\langle m \rangle)^{-1}\simeq
(1-\langle m \rangle)$. Since the value of $\langle m \rangle$ depends on galaxy selection for a given shape catalogue, 
we need to prepare the mock catalogues with an appropriate $\langle m \rangle$ on analysis-by-analysis basis in practice.


\section{CONCLUSIONS}
\label{sec:conclusions}

In this paper, we have presented 
a set of mock catalogues of galaxy shapes 
in the first-year
Subaru Hyper Suprime-Cam (HSC) data, referred to as HSC S16A.
We produced the mock catalogues constructed from full-sky gravitational lensing simulations in \citet{2017ApJ...850...24T}
and properly incorporated them with the observed galaxy information, 
allowing us to include various non-trivial effects in our mock catalogues.
By construction, our mock catalogues can include inhomogeneous angular distribution of source galaxies,
statistical uncertainty in photometric redshift (photo-$z$) estimate of each galaxy,
variations in the lensing weights due to observational conditions,
and the noise in galaxy shape measurement induced by both of intrinsic shape and measurement uncertainty. 
Using 2268 realizations of our mock catalogues in the footprint of HSC S16A, 
we performed a realistic analysis of galaxy-shape auto correlation functions $\xi_{\pm}$  
with tomographic redshift information
and predicted one of the most important statistical properties in the current cosmological parameter inference, the covariance matrices of $\xi_{\pm}$.
We compared the statistical property of $\xi_{\pm}$ in our mock catalogues with a theoretical model 
including the effect of nonlinear evolution in the matter density perturbations over cosmic time.
Furthermore, we studied several effects on the cosmological analyses of cosmic shear correlation functions,
including field variations among separated observed footprints,
photo-$z$ uncertainty of individual source galaxies,
and non-Gaussianity in the likelihood function.
Our findings are summarized as follows:

\begin{enumerate}

\vspace{2mm}
\item The ensemble average of shape correlation functions $\xi_{\pm}$ in our mock catalogues are 
consistent with a theoretical prediction based on the fitting formula 
of non-linear matter power spectrum \citep{Takahashi2012} within a $<10\%$ level accuracy.
The discrepancy from theoretical predictions can be explained by the finite angular resolution and sampling effect in
comoving distance in the ray-tracing simulations in \citet{2017ApJ...850...24T}.
Once the simulation-related effects are properly included, the ensemble average of $\xi_{\pm}$ over our mocks
is in agreement with its expectation with a level of $2-3\%$.

\vspace{2mm}
\item A theoretical model based on halo-model approach \citep[e.g.][]{2001ApJ...554...56C, 2009MNRAS.395.2065T, 2013PhRvD..87l3504T} is in good agreement with the mock covariance of $\xi_{\pm}$ at angular separations less than 10 arcmin,
whereas some corrections are required for 
the Gaussian covariance prediction in \citet{2008A&A...477...43J}.
The corrections arise from finite sampling of source galaxies in a limited sky coverage.
Once the corrections are included, the Gaussian covariance 
can provide a good fit to the mock covariance at degree scales.
We also found the non-Gaussian covariance coming from the mode coupling between super-survey and sub-survey modes
are dominant in the mock covariance at $\theta<10$ arcmin, while another term from the four-point correlations 
within a given survey window is less important \citep[also, see][]{2018arXiv180704266B}.

\vspace{2mm}
\item
We studied the effect of field variance among six separated patches in HSC S16A on the covariance of $\xi_{\pm}$.
Based on 2268 realizations, we found a $3-5\%$ level difference in the mock variance when we removed the effect 
of field variations on an estimator of covariance (see Figure~\ref{fig:HSCS16A_xis_var_fvimpact}). 
Since the Gaussian uncertainty in the variance is estimated to be $\sqrt{2/2268}\sim2.9\%$, we concluded that 
the field variation in the current HSC survey is less important for the covariance estimation.

\vspace{2mm}
\item
In addition to our fiducial photo-$z$ estimate, we also considered three different photo-$z$ estimates obtained from
different photo-$z$ pipelines and applied them to the mock catalogue production.
We then compared the signal-to-noise ratio of $\xi_{\pm}$ at a given angular separation
over four different photo-$z$ estimates for a fixed source selection. 
We found a $\sim5\%$ difference in the signal-to-noise ratio depending on the method of photo-$z$ estimation 
(see Figure~\ref{fig:HSCS16A_XMM_xis_var_diff_pz}). 
Our results show the systematic uncertainty due to photo-$z$ estimation
will be a minor issue in the covariance estimation of $\xi_{\pm}$ in HSC S16A data,
if one adopt the photo-$z$ estimates for the estimation of expectation value and their covariance in a self-consistent way.

\vspace{2mm}
\item
We predicted the total (cumulative) signal-to-noise ratio of cosmic shear correlation functions
in an angular range of $\theta=10-100$ arcmin with four tomographic bins based on our mock catalogues.
We found the expected signal-to-noise ratio to be $\sim16-17$ if our fiducial cosmological model would be 
correct and there are no multiplicative biases. 
A $\sim6\%$ difference has been found in our mock catalogues when we ignore the field variance 
among six separated patches in HSC S16A, while the difference is still comparable to the scatter in the evaluation 
of signal-to-noise ratio based on bootstrap sampling.

\vspace{2mm}
\item
Cosmological parameter inference with cosmic shear correlation relies on the chi-squared quantity defined as 
Eq.~(\ref{eq:chisq}). When data of interest follows Gaussian, the chi-squared quantity should 
follow the chi-squared distribution with degrees of freedom being the number of data bins. 
We validated if the chi-squared quantity for $\xi_{\pm}$ in HSC S16A can follow the expected distribution.
We found the variance of the chi-squared quantity evaluated from 2268 mock realizations are 1.42 times 
as large as the simple expectation based on the number of data bins. In addition,
the lower limit of 95\% confidence interval in the chi-squared quantity in our mock moves downward by $\sim$10\% 
compared to the expectation from the relevant chi-squared distribution (see Figure~\ref{fig:HSCS16A_xis_chi_sq_dist}).
We also found the degree-scale correlation functions induce the deviation from Gaussianity in likelihood function.

\vspace{2mm}
\item
We proposed a simple method to include non-zero multiplicative biases $\langle m\rangle$ 
in mock shape catalogues while keeping the shear responsivity fixed.
Applying this method to tomographic cosmic shear correlation analysis in the HSC S16A, we examined the impact
of multiplicative biases on the covariance. We found that the effective shape noise term can change by a factor of $(1+\langle m\rangle)^{-1}$.
Hence, the small-scale covariance of cosmic shear correlation functions can be affected by the presence of non-zero $\langle m\rangle$.
For the HSC S16A cosmic shear correlation analysis with four tomographic redshift bins, we found the cumulative signal-to-noise ratio
can degrade by $\simeq20\%$ level when including non-zero $\langle m\rangle$ to estimate the covariance.

\end{enumerate}

Since our mock catalogues take into account a lot of relevant features in galaxy shape measurement, 
one can utilize them for various purposes. 
Those include the covariance estimation of {\it any} cosmic shear statistics \citep[e.g.][for a recent review]{2015RPPh...78h6901K},
and validation of analysis pipeline for parameter estimation \citep[e.g. see][for the representative work]{2017arXiv170609359K}.
In spite of our effort making the mock catalogues realistic as possible, there still remain some room for developing
more realistic mock catalogues of galaxy shapes. An important feature missed in our mock catalogues is the correlation 
between property of source galaxy and matter density distribution at the redshift which the source galaxy locates. Intrinsic alignment (IA) effect of galaxy shape is among the most typical effects \citep{2004PhRvD..70f3526H, 2015PhR...558....1T} and future work should include the IA effect properly.
Also, source-lens clustering effects \citep{2002MNRAS.330..365H} and magnification effects on observed galaxy \citep{2009PhRvL.103e1301S, 2014PhRvD..89b3515L} can be an issue in mock catalogue production for future lensing surveys.

Mock catalogue for modern galaxy survey should be also available for covariance estimation of cross correlation 
analyses of galaxy shapes with large-scale structures. For this purpose, one needs to construct a synthetic catalogue of tracers of large-scale structures and preserve the statistical correlation between galaxy shapes in mock catalogue productions.
Since our mock shape catalogues are based on full-sky lensing simulations and their inherent halo catalogues are also available,
we can construct mock catalogues of foreground objects for cross correlation with HSC S16A galaxy shapes
by using a similar technique as in \citet{2017MNRAS.470.3476S}. We leave it for our future work 
to create the mock catalogues of tracers of large-scale structures, that are correlated with mock shape catalogues 
in this paper.


\section*{acknowledgments} 

This work is in part supported by MEXT Grant-in-Aid for Scientific Research on Innovative Areas (No.~15H05887, 15H05893, 15K21733),
by MEXT KAKENHI Grant Number (15H03654, 18H04358),
and by JSPS Program for Advancing
Strategic International Networks to Accelerate the Circulation of Talented Researchers.

The Hyper Suprime-Cam (HSC) collaboration includes the astronomical communities of Japan and Taiwan, and Princeton University.  The HSC instrumentation and software were developed by the National Astronomical Observatory of Japan (NAOJ), the Kavli Institute for the Physics and Mathematics of the Universe (Kavli IPMU), the University of Tokyo, the High Energy Accelerator Research Organization (KEK), the Academia Sinica Institute for Astronomy and Astrophysics in Taiwan (ASIAA), and Princeton University.  Funding was contributed by the FIRST program from Japanese Cabinet Office, the Ministry of Education, Culture, Sports, Science and Technology (MEXT), the Japan Society for the Promotion of Science (JSPS),  Japan Science and Technology Agency  (JST),  the Toray Science  Foundation, NAOJ, Kavli IPMU, KEK, ASIAA,  and Princeton University.

The Pan-STARRS1 Surveys (PS1) have been made possible through contributions of the Institute for Astronomy, the University of Hawaii, the Pan-STARRS Project Office, the Max-Planck Society and its participating institutes, the Max Planck Institute for Astronomy, Heidelberg and the Max Planck Institute for Extraterrestrial Physics, Garching, The Johns Hopkins University, Durham University, the University of Edinburgh, Queen's University Belfast, the Harvard-Smithsonian Center for Astrophysics, the Las Cumbres Observatory Global Telescope Network Incorporated, the National Central University of Taiwan, the Space Telescope Science Institute, the National Aeronautics and Space Administration under Grant No. NNX08AR22G issued through the Planetary Science Division of the NASA Science Mission Directorate, the National Science Foundation under Grant No. AST-1238877, the University of Maryland, and Eotvos Lorand University (ELTE).

This paper makes use of software developed for the Large Synoptic Survey Telescope. We thank the LSST Project for making their code available as free software at \verb|http://dm.lsst.org|.

Based [in part] on data collected at the Subaru Telescope and retrieved from the HSC data archive system, which is operated by the Subaru Telescope and Astronomy Data Center at National Astronomical Observatory of Japan.

Numerical computations were in part carried out on Cray XC30 and XC50 at Center for Computational Astrophysics, National Astronomical Observatory of Japan, and on XC40 at YITP in Kyoto University.


\bibliographystyle{mnras}
\bibliography{bibtex_library}

\appendix
\section{Finite area effect of Gaussian covariance in cosmic shear tomography}
\label{apdx:finite_area_effect}

As shown in \citet{Sato2011},
the estimator of cosmic shear correlation given by Eq.~(\ref{eq:xi_pm_est_tomo})
does not always have a predicted covariance (Eq.~\ref{eq:cov_G_xi_predict}) 
even if shear field is assumed to be Gaussian. Eq.~(\ref{eq:cov_G_xi_predict})
will be valid only for wide-area surveys with sky coverage of 
$\simgt1000$ squared degrees, whereas one need to evaluate the Gaussian covariance
from more direct expressions of covariance of estimator itself
\citep[see, Eqs.(23)-(25) in][]{2002A&A...396....1S}
when the sky coverage is of an order of 100 squared degrees.
We here extend the formula in \citet{2002A&A...396....1S} 
by considering the correlation of source galaxies at different redshifts.

Let us return to the definition of the covariance of Eq.~(\ref{eq:xi_pm_est_tomo}):
\begin{align}
&{\rm Cov}\left[\hat{\xi}_{\pm,ab}(\theta_{1}), \hat{\xi}_{\pm, cd}(\theta_{2})\right] \nonumber \\
&\,\,\,\,\,\,\,\,\,\,\,\,\,\,\,\,\,\,\,
\equiv \langle (\hat{\xi}_{\pm,ab}(\theta_{1})-\xi_{\pm,ab}(\theta_{1})) (\hat{\xi}_{\pm,cd}(\theta_{2})-\xi_{\pm,cd}(\theta_{2}))\rangle. \label{eq:def_cov_xi_tomo}
\end{align}
In the calculation of Eq.~(\ref{eq:def_cov_xi_tomo}), 
the four point correlation function of distortion appears.
Assuming that the shear field and the source distortion are Gaussian and ignoring the shape noise terms\footnote{The finite area effect becomes important only for large scales comparable to the size of survey window,
making this assumption valid.
}, 
we can write the four point correlation as the product of the two point function as follows; 
\begin{align}
&\frac{\langle \epsilon^{(a)}_{\alpha i}\epsilon^{(b)}_{\beta j}
\epsilon^{(c)}_{\mu k}\epsilon^{(d)}_{\nu \ell}\rangle}{(2{\cal R}^{(a)})(2{\cal R}^{(b)})(2{\cal R}^{(c)})(2{\cal R}^{(d)})}
= 
\langle \gamma^{(a)}_{\alpha i}\gamma^{(b)}_{\beta j}\rangle \langle \gamma^{(c)}_{\mu k}\gamma^{(d)}_{\nu \ell}\rangle
\nonumber \\
&\,\,\,\,\,\,\,\,\,\,\,\,\,\,\,\,\,\,\,\,\,\,\,\,\,
+\langle \gamma^{(a)}_{\alpha i}\gamma^{(c)}_{\mu k}\rangle \langle \gamma^{(b)}_{\beta j}\gamma^{(d)}_{\nu \ell}\rangle
+\langle \gamma^{(a)}_{\alpha i}\gamma^{(d)}_{\nu \ell}\rangle \langle \gamma^{(b)}_{\beta j}\gamma^{c}_{\mu k}\rangle,\label{eq:eeee}
\end{align}
where the above equation is valid for $i \ne j$ and $k \ne \ell$ and Greek letters represent 1 or 2.

From Eq.~(\ref{eq:eeee}) and the fact that
\begin{align}
\epsilon^{(a)}_{ti}\epsilon^{(b)}_{tj}
+\epsilon^{(a)}_{\times i}\epsilon^{(b)}_{\times j} 
=& 
\epsilon^{(a)}_{1i}\epsilon^{(b)}_{1j}+\epsilon^{(a)}_{2i}\epsilon^{(b)}_{2j}, \\
\epsilon^{(a)}_{ti}\epsilon^{(b)}_{tj}
-\epsilon^{(a)}_{\times i}\epsilon^{(b)}_{\times j} 
=& (\epsilon^{(a)}_{1i}\epsilon^{(b)}_{1j}-\epsilon^{(a)}_{2i}\epsilon^{(b)}_{2j})\cos4\phi_{ij} \nonumber \\
&\,\,\,\,\,\,\,\,\,\,\,\,\,\,\,\,\,\,\,
+ (\epsilon^{(a)}_{1i}\epsilon^{(b)}_{1j}+\epsilon^{(a)}_{2i}\epsilon^{(b)}_{2j})\sin4\phi_{ij}, 
\end{align}
where $\phi_{ij}$ is the polar angle of 
$\bd{\theta}_{i}-\bd{\theta}_{j}$, 
we can express the covariance of Eq.~(\ref{eq:def_cov_xi_tomo}) 
as follows;
\begin{align}
&{\rm Cov}_{\rm G}\left[\hat{\xi}_{+,ab}(\theta_{1}),  \hat{\xi}_{+,cd}(\theta_{2})\right] =
\frac{1}{{\scriptstyle N^{ab}_{p}(\theta_{1})N^{cd}_{p}(\theta_2)}}
\sum_{ijk\ell}w^{(a)}_{i} w^{(b)}_{j} w^{(c)}_{k} w^{(d)}_{\ell} 
\nonumber \\
&\times\Delta_{\theta_{1}}(ij)\Delta_{\theta_{2}}(k\ell) 
\, \Bigl\{
\xi_{+,ac}(ik)\xi_{+,bd}(j\ell)
+\xi_{+,ad}(i\ell)\xi_{+,bc}(jk)
\nonumber \\
&
\,\,\,\,\,\,\,\,\,\,\,\,\,\,\,\,\,\,\,\,\,
\,\,\,
\,\,\,\,\,\,\,\,\,\,\,\,\,\,\,\,\,\,\,\,\,\,\,\,
+
\cos\left[4(\phi_{ik}-\phi_{j\ell})\right]
\xi_{-,ac}(ik)\xi_{-,bd}(j\ell)
\nonumber \\
&
\,\,\,\,\,\,\,\,\,\,\,\,\,\,\,\,\,\,\,\,\,
\,\,\,
\,\,\,\,\,\,\,\,\,\,\,\,\,\,\,\,\,\,\,\,\,\,\,\,
+
\cos\left[4(\phi_{i\ell}-\phi_{jk})\right]\xi_{-,ad}(i\ell)\xi_{-,bc}(jk)\Bigr\}/2, \label{eq:cov_pppp} \\
&{\rm Cov}_{\rm G}\left[\hat{\xi}_{-,ab}(\theta_{1}),  \hat{\xi}_{-,cd}(\theta_{2})\right]
=\frac{1}{{\scriptstyle N^{ab}_{p}(\theta_{1})N^{cd}_{p}(\theta_2)}}
\sum_{ijk\ell}w^{(a)}_{i} w^{(b)}_{j} w^{(c)}_{k} w^{(d)}_{\ell}
\nonumber \\
&\times
\Delta_{\theta_{1}}(ij)\Delta_{\theta_{2}}(k\ell)
\Bigl\{
\cos\left[4(\phi_{ij}+\phi_{k\ell}-\phi_{ik}-\phi_{j\ell})\right]
\xi_{-,ac}(ik)\xi_{-,bd}(j\ell)
\nonumber \\
&\,\,\,\,\,\,\,\,\,\,\,\,\,\,\,\,\,\,\,\,\,
+
\cos\left[4(\phi_{ij}+\phi_{k\ell}-\phi_{i\ell}-\phi_{jk})\right]
\xi_{-,ad}(i\ell)\xi_{-,bc}(jk)
\nonumber \\
&\,\,\,\,\,\,
+
\cos\left[4(\phi_{ij}-\phi_{k\ell})\right]
\left(\xi_{+,ac}(ik)\xi_{+,bd}(j\ell)+\xi_{+,ad}(i\ell)\xi_{+,bc}(jk)\right)\Bigr\}/2, 
\label{eq:cov_mmmm} \\
&{\rm Cov}_{\rm G}\left[\hat{\xi}_{+,ab}(\theta_{1}),  \hat{\xi}_{-,cd}(\theta_{2})\right]
=\frac{1}{{\scriptstyle N^{ab}_{p}(\theta_{1})N^{cd}_{p}(\theta_2)}}
\sum_{ijk\ell}w^{(a)}_{i} w^{(b)}_{j} w^{(c)}_{k} w^{(d)}_{\ell}
\nonumber \\
&\times
\Delta_{\theta_{1}}(ij)\Delta_{\theta_{2}}(k\ell)
\Bigl\{
\cos\left[4(\phi_{k\ell}-\phi_{j\ell})\right]
\xi_{+,ac}(ik)\xi_{-,bd}(j\ell)
\nonumber \\
&
\,\,\,\,\,\,\,\,\,\,\,\,\,\,\,\,\,\,\,\,\,\,\,\,
\,\,\,\,\,\,\,\,\,\,\,\,\,\,\,\,\,\,\,\,\,\,\,\,
+
\cos\left[4(\phi_{k\ell}-\phi_{ik})\right]
\xi_{-,ac}(ik)\xi_{+,bd}(j\ell) \nonumber \\
&
\,\,\,\,\,\,\,\,\,\,\,\,\,\,\,\,\,\,\,\,\,\,\,\,
\,\,\,\,\,\,\,\,\,\,\,\,\,\,\,\,\,\,\,\,\,\,\,\,
+
\cos\left[4(\phi_{k\ell}-\phi_{jk})\right]
\xi_{+,ad}(i\ell)\xi_{-,bc}(jk)
\nonumber \\
&
\,\,\,\,\,\,\,\,\,\,\,\,\,\,\,\,\,\,\,\,\,\,\,\,
\,\,\,\,\,\,\,\,\,\,\,\,\,\,\,\,\,\,\,\,\,\,\,\,
+
\cos\left[4(\phi_{k\ell}-\phi_{i\ell})\right]
\xi_{-,ad}(i\ell)\xi_{+,bc}(jk)\Bigr\}/2, \label{eq:cov_ppmm}
\end{align}
where $\xi_{\pm,ab}(ik) = \xi_{\pm,ab}(|\bd{\theta}_{i}-\bd{\theta}_{k}|)$ and so on.

\section{The finite thickness effect in ray-tracing simulations}
\label{app:finite_thickness}

In this appendix, we summarize the effect of finite sampling in comoving distances in 
multiple-plane ray-tracing simulations.
Here we suppose that the ray-tracing simulations have been constructed by $N$ shells of projected mass density at different redshifts. In this case, the integral in Eq.~(\ref{eq:kappa_delta}) in the ray-tracing simulations should be expressed as
\begin{align}
\kappa_{\rm sim}({\bd \theta}) =& \sum_{i=1}^{N} \Delta \chi \, q(\chi_i) \, \delta^{\rm shell}({\bd \theta}, \chi_i) \label{eq:kappa_delta_sim} \\
\delta^{\rm shell}({\bd \theta}, \chi_i) =& \frac{1}{\chi^{2}_{i}\Delta \chi}\, 
\int_{\chi_{i}-\Delta \chi/2}^{\chi_{i}+\Delta \chi/2}\, {\rm d}\chi^{\prime}(\chi^{\prime})^2\, \delta_{\rm m}(\chi^{\prime}, r(\chi^{\prime}){\bd \theta}), \label{eq:density_shell}
\end{align}
where $\chi_{i} = (i-1/2)\Delta \chi$ and we set $N=38$ and $\Delta \chi = 150\, h^{-1}\, {\rm Mpc}$ in our case.
Under the Born approximation, 
the lensing power spectrum in the ray-tracing simulation can be computed as 
the summation of the power spectrum of density shells, denoted as $P^{W}_{\rm m}$.
The analytic expression of $P^{W}_{\rm m}$ is found in Appendix B in \citet{2017ApJ...850...24T}.
In \citet{2017ApJ...850...24T}, the authors also provide a simple approximated formula of $P^{W}_{\rm m}$ 
at $i$-th density shell as 
\beq
P^{W}_{\rm m}(k, z(\chi_i)) = \frac{(1+c_1 k^{-\alpha_1})^{\alpha_1}}{(1+c_2 k^{-\alpha_2})^{\alpha_3}}\, P_{\rm m}(k, z),
\label{eq:P_m_shell}
\eeq
with
$c_1 = 9.5171\times 10^{-4}$,
$c_2 = 5.1543\times 10^{-3}$,
$\alpha_1=1.3063$,
$\alpha_2=1.1475$, 
and
$\alpha_3= 0.62793$.
Note that the wavenumber $k$ is in unit of $h \,{\rm Mpc}^{-1}$ and the correction term in Eq.~(\ref{eq:P_m_shell}) is independent of redshift.
In addition, the finite resolution in comoving distance in the simulation should be included in 
the computation of lensing kernel (Eq.~\ref{eq:lens_kernel}).
We estimate the coarse grained $p(z)$ in the simulation by degrading the original $p(z)$ as shown in Figure~\ref{fig:HSCS16A_pz_4bins} with redshift.  
In summary, we model the finite thickness effect in the lensing power spectrum as
\begin{align}
P_{\kappa {\rm sim}, ab}(\ell) 
=& \sum_{i=1}^{N} \frac{\Delta \chi}{r^{2}_{\rm eff,i}} \, 
q_a(\chi_i) \, q_b(\chi_i)\, P^{W}_{\rm m}\left(\frac{\ell}{r_{\rm eff,i}}, z(\chi_i)\right), \label{eq:Pkappa_sim_finitez}\\
r_{\rm eff,i} =& \frac{3}{4}\frac{(r^{4}_{2,i}-r^{4}_{1,i})}{(r^{3}_{2,i}-r^{3}_{1,i})},
\end{align}
where $r_{\rm eff,i}$ is the cone-weighted comoving distance with 
$r_{2,i}=r(\chi_i+\Delta\chi/2)$ and $r_{1,i}=r(\chi_i+\Delta\chi/2)$ \citep[see also][]{2015MNRAS.453.3043S}.

\section{A halo model for covariance estimation of cosmic shear}
\label{apdx:halo_model}

In this appendix, we summarize the formulation based on halo-model approach that used to predict
the cosmic shear covariance. We follow the method as in \citet{2001ApJ...554...56C}, \citet{2009MNRAS.395.2065T} and \citet{2013PhRvD..87l3504T}.

The covariance of cosmic shear tomographic analyses can be decomposed into three terms as shown in Eq.~(\ref{eq:cov_decomp}).
Among these, the Gaussian covariance can be computed with combinations of cosmic shear power spectra.
In computing of power spectrum for a given set of tomographic bins (see Eq.~\ref{eq:P_kappa}), 
we adopt the fitting formula of the non-linear matter power spectrum developed in \citet{Takahashi2012}.

On the non-Gaussian covariance, we require a theoretical model of weak lensing trispectrum.
Weak lensing trispectrum is defined as (with respect to convergence $\kappa$)
\beq
\langle \tilde{\kappa}_{a}({\bd \ell}_1) \tilde{\kappa}_{b}({\bd \ell}_2) \tilde{\kappa}_{c}({\bd \ell}_3) \tilde{\kappa}_{d}({\bd \ell}_4)\rangle
\equiv (2\pi)^2 \delta^{(2)}({\bd \ell}_{1234}) T_{\kappa, abcd}({\bd \ell}_1, {\bd \ell}_2, {\bd \ell}_3, {\bd \ell}_4), \label{eq:def_Tkappa}
\eeq
where ${\bd \ell}_{ij\cdots n} = {\bd \ell}_{i}+{\bd \ell}_{j}+\cdots+{\bd \ell}_{n}$. 
Under the Limber approximation, the trispectrum can be computed as
\begin{align}
T_{\kappa, abcd}({\bd \ell}_1, {\bd \ell}_2, {\bd \ell}_3, {\bd \ell}_4) =& 
\int_{0}^{\chi_H}\, \frac{\rm d\chi}{r^6(\chi)} q_a(\chi)\, q_b(\chi)\, q_c(\chi)\, q_d(\chi) \nonumber \\
&\,\,\,\,\,\,\,\,\,\,\,\,\,\,\,\,\,\,\,\,\,
\times
T_{\rm m}({\bd k}_1, {\bd k}_2, {\bd k}_3, {\bd k}_4, z(\chi)),
\end{align}
where ${\bd k}_{i} = {\bd \ell}_{i}/\chi$
and $T_{\rm m}$ represents the trispectrum of matter overdensity field as defined in a similar way to Eq.~(\ref{eq:def_Tkappa}).
Previous studies have shown that the dominant contribution of the non-Gaussian covariance in cosmic shear at relevant scales of $\ell\simgt100$ is 
given by the so-called one-halo term and the SSC terms \citep[e.g.][]{Sato2009}.
The one-halo term arises from the four point correlation among different Fourier modes in single dark matter halos and it is expressed as
\beq
T^{1h}_{\rm m}({\bd k}_1, {\bd k}_2, {\bd k}_3, {\bd k}_4, z(\chi))= 
\int {\rm d}M\, \frac{{\rm d}n}{{\rm d}M}\left(\frac{M}{\bar{\rho}_{\rm m}}\right)^4\, \tilde{u}_{1}\,\tilde{u}_{2}\,\tilde{u}_{3}\,\tilde{u}_{4}, 
\eeq
where ${\rm d}n/{\rm d}M$ is the halo mass function,
$\tilde{u}$ is the Fourier counterpart of normalized halo density profile 
(the normalization is set so that the volume integral of the density profile should be unity),
and $\tilde{u}_{i} = \tilde{u}({\bd k}_{i}, z, M)$.
To compute the term of $\tilde{u}({\bd k}_{i}, z, M)$, we adopt the NFW profile \citep{Navarro1996}
with halo concentration as in \citet{2015ApJ...799..108D}.
For the halo mass function, we adopt the fitting formula developed in \citet{2008ApJ...688..709T} throughout this paper.

Another important contributor to the non-Gaussian covariance is the SSC term which includes the four point correlation
among super-survey modes. As shown in \citet{2013PhRvD..87l3504T}, the SSC term can be given by Eqs.~(\ref{eq:T_SSC}) and (\ref{eq:sigma_W}).
To compute the SSC term, we adopt the halo model of the response of matter power spectrum as follows \citep{2013PhRvD..87l3504T}
\begin{align}
&\frac{\partial P_{\rm m}(k, z)}{\partial \delta_{\rm bg}} = I^{1}_{2}(k,z) \nonumber \\
&+
\left(\frac{68}{21}-\frac{1}{3}\frac{{\rm d}\ln k^3\, \left[I^{1}_{1}(k,z)\right]^2  P_{\rm L}(k,z)}{{\rm d}\ln k}\right)
\left[I^{1}_{1}(k,z)\right]^2 P_{\rm L}(k,z),
\end{align}
where we use the notation as in \citet{2001ApJ...554...56C}:
\beq
I^{\beta}_{\mu}(k_1, k_2,\cdots, k_{\mu}) \equiv \int {\rm d}M\, \frac{{\rm d}n}{{\rm d}M}\, b_{\beta}\, \tilde{u}_{1}\,\tilde{u}_{2}\cdots\tilde{u}_{\mu}, 
\eeq
where $b_{0}=1$ and $b_1$ is set to be the linear halo bias. In this paper, we apply the model of linear halo bias in \citet{2010ApJ...724..878T}.

It is worth noting that the non-Gaussian covariance from the SSC trispectrum requires the computation of the variance in matter density for a given survey window. In this paper, we properly include the mask in real HSC S16A to compute Eq.~(\ref{eq:sigma_W}).
To do so, we first generate a Gaussian density field on a flat sky by using the linear power spectrum at redshift of interest $z$.
When generating random Gaussian density field, we set the sky coverage to be $20\times20$ squared degrees
and pixel size to be $0.3$ arcmin. Then, we paste the survey window subtracted from real HSC S16A onto a squared sky. 
The survey window of 6 different HSC S16A patches 
is defined as in Section~\ref{sec:data}.
We confirm that the field of view on a flat sky is large enough to cover the whole survey window in individual HSC S16A fields
and the pixel size is sufficiently small to trace a complex geometry of the survey window.
After pasting the mask, we compute the variance of Gaussian density field within the survey window.
We repeat the above procedures ten times to reduce the scatter in the estimation of $\sigma_W(z)$.
The variance estimation has been performed at discrete 20 points in redshifts between $z=0.001$ and $10$
with logarithmic binning of $\Delta \log z = 0.2$. When computing the SSC term in different HSC S16A fields, 
we interpolate the precomputed 20 data of $\sigma_W(z)$.

\section{An estimator of field variation in cosmic shear two-point correlation functions}
\label{app:est_fv}

\begin{figure}
\centering
\includegraphics[width=0.95\columnwidth]
{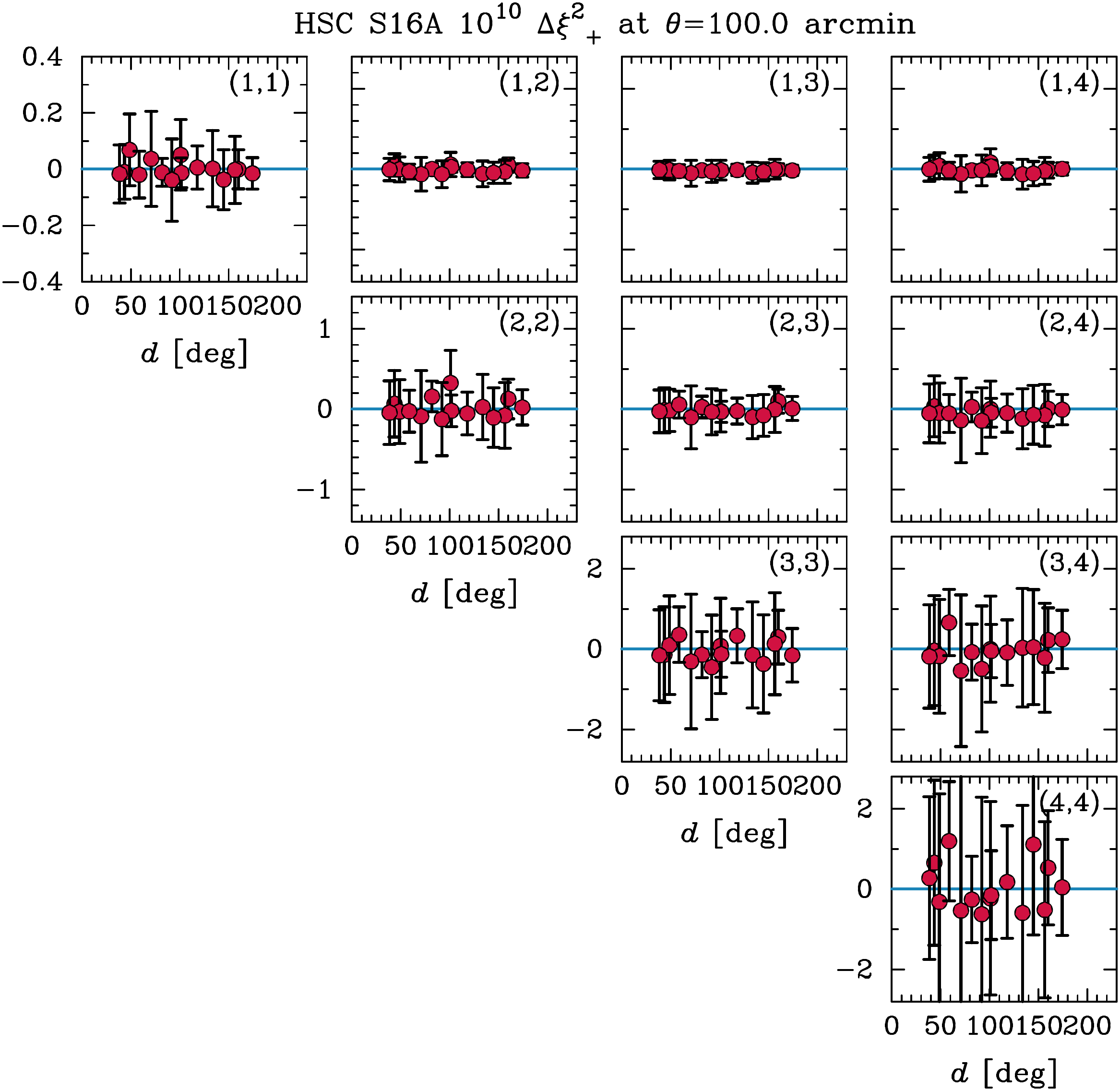}
\includegraphics[width=0.95\columnwidth]
{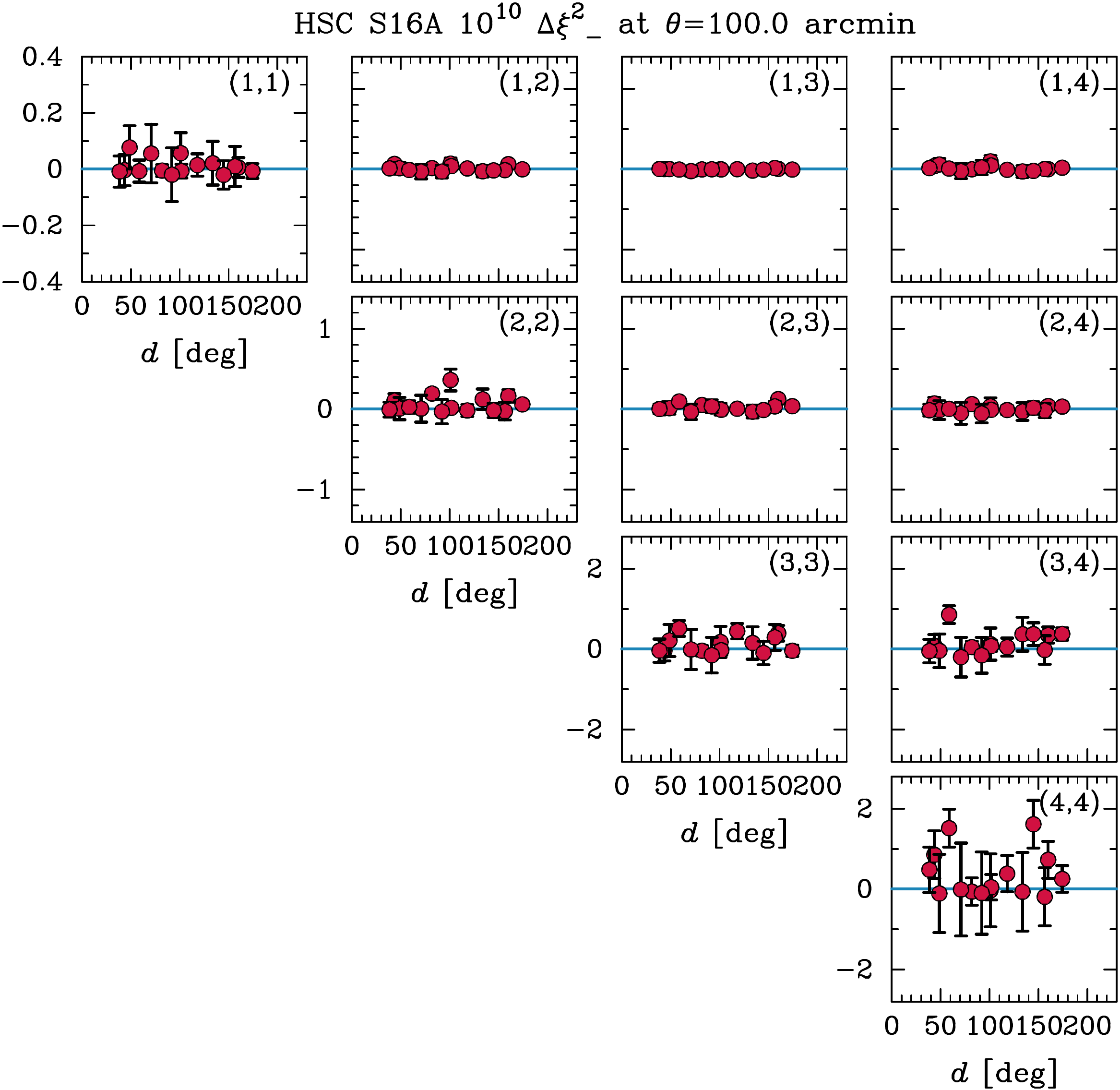}
\caption{
The field variation of $\xi_{\pm}$ in HSC S16A.
We here show the square of difference of $\xi_{\pm}$ between two separated HSC S16A patches
as a function of separation length between two patches $d$.
The estimator is constructed so that we can have a null signal if the measurements of $\xi_{\pm}$ 
in two patches are independent each other. The detail of the estimator is found in Eq.~(\ref{eq:est_xi_fv}).
The top panels show the results for $\xi_{+}$, while the bottom is for $\xi_{-}$.
In both, 10 small panels represents the results with different set of tomographic bins.
The legend $(1,2)$ represents the cosmic shear correlation analysis with first and second tomographic bins,
and so on. In each panel, red points are from the HSC S16A shape catalogue and we evaluate error bars 
from 2268 mock realizations.
}
\label{fig:HSCS16A_shear_xis_diff_sq}
\end{figure}

Since our mock catalogues of HSC S16A shape preserve a proper positional relationship among six separated patches,
those allow us to evaluate the field variation on cosmic shear correlation functions $\xi_{\pm}$ in HSC S16A.

We here consider the cosmic shear analyses with four different photometric bins as worked in the main text.
Suppose that the measurement of $\xi_{\pm}$ is carried out on individual separated patches, 
we will characterize a field variation on the measured $\xi_{\pm}$ among different patches by
\begin{align}
&\Delta \xi_{\pm}^2 ({\rm field}\, 1 |{\rm field}\,2) 
\equiv 
\left[\xi_{\pm}({\rm field}\, 1) - \xi_{\pm}({\rm field}\, 2)\right]^2 \nonumber \\
&
\,\,\,\,\,\,\,\,\,\, \,\,\,\,\,\,\,\,\,\, \,\,\,\,\,\,\,\,\,\,
\,\,\,\,\,\,\,\,\,\, \,\,\,\,\,\,\,\,\,\, 
- {\rm Var}\left[\xi_{\pm}({\rm field}\, 1)\right]
- {\rm Var}\left[\xi_{\pm}({\rm field}\, 2)\right] , \label{eq:est_xi_fv}
\end{align}
where $\xi_{\pm}({\rm field}\, i)$ represents the observed cosmic shear correlation function on $i$-th field,
and ${\rm Var}\left[\xi_{\pm}({\rm field}\, i)\right]$ is its variance.
The variances in Eq.~(\ref{eq:est_xi_fv}) can be directly estimated from our mock catalogues.
In addition, our mock catalogues enable us to set the statistical uncertainty of Eq.~(\ref{eq:est_xi_fv})
when we apply the estimator to 2268 realizations of mock catalogues. Note that Eq.~(\ref{eq:est_xi_fv}) will be evaluated for a given angular separation in $\xi_{\pm}$ and set of tomographic bins.
We construct Eq.~(\ref{eq:est_xi_fv}) so that we will have a null signal on average 
($\langle \Delta \xi_{\pm}^2 ({\rm field}\, 1 |{\rm field}\, 2|)\rangle =0$)
if the measurement of $\xi_{\pm}$ is independent of a choice of survey patches.
In this appendix, we study Eq.~(\ref{eq:est_xi_fv}) as a function of separation length between two fields $d$.

Figure~\ref{fig:HSCS16A_shear_xis_diff_sq} summarizes the results of Eq.~(\ref{eq:est_xi_fv}) when we apply 
to the HSC S16A shape catalogue. In this figure, we work with $\xi_{\pm}$ at 100 arcmin and scale the correlation function
by a factor of $10^5$. The top and bottom panels represent the results for $\xi_{+}$ and $\xi_{-}$, respectively.
Each small panel in the top and bottom shows difference in tomographic bins in the analysis.
In this figure, the red point is the actual measurement of Eq.~(\ref{eq:est_xi_fv}) in the HSC S16A data
and error bars are estimated from 2268 mock realizations. As seen in the figure, there are 
no clear trends of field variation of $\xi_{\pm}$ as a function of separation length between separated fields,
whereas detailed analyses will be interesting to measure super-sample density fluctuations whose wavelengths are 
larger than the survey scale in a more direct way \citep[e.g.][]{2014PhRvD..90j3530L}.
We leave it for our future work.

\end{document}